\documentclass[12pt]{article}
\usepackage[T1]{fontenc}
\usepackage[margin=1in]{geometry}

\usepackage{amsmath}
\usepackage{accents, adjustbox, afterpage, amssymb, amsfonts, amsthm, array, booktabs, calc, caption, color, comment, datetime, dcolumn, enumitem, eurosym, float, geometry, graphics, graphicx, longtable, mathtools, mathbbol, multirow, natbib, pdflscape, pdfpages, ragged2e, refcount, rotating, sectsty, setspace, subcaption, subfiles, textpos, threeparttable, tikz, soul, xcolor, xparse
}
\DeclareMathOperator*{\argmax}{arg\,max}

\usepackage[hang, flushmargin]{footmisc}

\usepackage[ruled,vlined]{algorithm2e}
\usepackage{hyperref}
\usepackage{bookmark}
\setlist[enumerate]{itemsep=0pt,topsep=2pt}
\AtBeginDocument{
 \setlength\abovedisplayskip{4pt}
 \setlength\belowdisplayskip{4pt}
}
\usepackage[sf,sl,outermarks]{titlesec}
\newcommand{\addperiod}[1]{#1.}

\titleformat{\section}[block]
{\normalfont\Large\bfseries}{\thesection.}{.5em}{\Large\bfseries}
\titlespacing*{\section}{0pt}{*1.3}{*0.2}

\titleformat{\subsection}[block]
{\normalfont\large\bfseries}{\thesubsection.}{.5em}{\large\bfseries}
\titlespacing*{\subsection}{0pt}{*1}{*0}

\titleformat{\subsubsection}[runin]
{\normalfont\bfseries}{}{0em}{\normalsize\bfseries\addperiod}
\titlespacing*{\subsubsection}{0pt}{*.15}{*1}

\NewDocumentCommand{\hyref}{m O{}O{}}{\hyperref[#1]{#2 \ref{#1}#3}}
\setcitestyle{authoryear,aysep={},open={(},close={)}}
\bibpunct{(}{)}{;}{a}{,}{,}

\usepackage{etoolbox}

\makeatletter
\DeclareRobustCommand\citepos													% Create possessives with \citepos[e.g.][p. Y]{Key}: e.g. Key's (Year, p. Y)
  {\begingroup\def\NAT@nmfmt##1{{\NAT@up##1's}}%
   \NAT@swafalse\let\NAT@ctype\z@\NAT@partrue
   \@ifstar{\NAT@fulltrue\NAT@citetp}{\NAT@fullfalse\NAT@citetp}}

\pretocmd{\NAT@citex}{%
  \let\NAT@hyper@\NAT@hyper@citex
  \def\NAT@postnote{#2}%
  \setcounter{NAT@total@cites}{0}%
  \setcounter{NAT@count@cites}{0}%
  \forcsvlist{\stepcounter{NAT@total@cites}\@gobble}{#3}}{}{}
\newcounter{NAT@total@cites}
\newcounter{NAT@count@cites}
\def\NAT@postnote{}

\def\NAT@hyper@citex#1{%														% Include postnote and \citet closing bracket in hyperlink
  \stepcounter{NAT@count@cites}%
  \hyper@natlinkstart{\@citeb\@extra@b@citeb}#1%
  \ifnumequal{\value{NAT@count@cites}}{\value{NAT@total@cites}}
    {\if*\NAT@postnote*\else\NAT@cmt\NAT@postnote\global\def\NAT@postnote{}\fi}{}%
  \ifNAT@swa\else\if\relax\NAT@date\relax
  \else\NAT@@close\global\let\NAT@nm\@empty\fi\fi								% Avoid compact citations
  \hyper@natlinkend}
\renewcommand\hyper@natlinkbreak[2]{#1}

\patchcmd{\NAT@citex}															% Avoid extraneous postnotes, closing brackets
  {\ifNAT@swa\else\if*#2*\else\NAT@cmt#2\fi
   \if\relax\NAT@date\relax\else\NAT@@close\fi\fi}{}{}{}
\patchcmd{\NAT@citex}
  {\if\relax\NAT@date\relax\NAT@def@citea\else\NAT@def@citea@close\fi}
  {\if\relax\NAT@date\relax\NAT@def@citea\else\NAT@def@citea@space\fi}{}{}
\patchcmd{\NAT@cite}{\if*#3*}{\if*\NAT@postnote*}{}{}
\makeatother

\usepackage{dsfont,fouriernc,charter,libertine}

\newtheoremstyle{newtheorem}{5pt}{2pt}{\itshape}{0pt}{\bfseries}{.}{.4em}{\thmname{#1}\thmnumber{ #2}\textnormal{\thmnote{ (#3)}}}
\theoremstyle{newtheorem}

\newtheorem{hypothesis}{Hypothesis}

\newtheorem{proc}{Procedure}

\definecolor{cblue}{rgb}{155, 221, 255}

\hypersetup{
 colorlinks,
 linkcolor={red!50!black},
 citecolor={blue!50!black},
 urlcolor={blue!50!black}
}

\newdateformat{DDMonthYYYY}{\THEDAY\, \monthname[\THEMONTH] \THEYEAR}

\definecolor{mygreen}{RGB}{0, 150, 0}
\definecolor{myred}{RGB}{207,8,8}

\hypersetup{
 pdffitwindow=false,colorlinks=true,linkcolor={red!50!black},citecolor={blue!50!black},urlcolor=black,linkbordercolor={white},
}

\begin{document}
\thispagestyle{empty}
\setcounter{page}{0}
\setcounter{footnote}{0}
\renewcommand{\thefootnote}{\fnsymbol{footnote}}
~\vspace*{-2cm}\\
\begin{center}
 {\noindent
 \Large
 \bfseries The Informational Role of Online Recommendations:\\[.5em] 
 Evidence from a Field Experiment\footnotemark
 }
\end{center}

\makebox[\textwidth][c]{
 \begin{minipage}{1.2\linewidth}
 \Large\centering
 \begin{tabular}{lr}
 Guy Aridor &
 \quad Duarte Gon\c{c}alves
 \end{tabular}
 \begin{tabular}{clr}
 Daniel Kluver &
 \quad Ruoyan Kong &
 \quad Joseph Konstan
 \end{tabular}
 \end{minipage}
}
\footnotetext{
 \setstretch{1}\emph{Contact:} 
 \setstretch{1} Aridor: Northwestern Kellogg, \hyperlink{mailto:guy.aridor@kellogg.northwestern.edu}{\color{black}guy.aridor@kellogg.northwestern.edu}. 
 Gon\c{c}alves: University College London; \hyperlink{mailto:duarte.goncalves@ucl.ac.uk}{\color{black}duarte.goncalves@ucl.ac.uk}. 
 Kluver and Kong and Konstan: GroupLens and University of Minnesota - Twin Cities; \hyperlink{mailto:kluve018@umn.edu}{\color{black}kluve018@umn.edu}, \hyperlink{mailto:kong0135@umn.edu}{\color{black}kong0135@umn.edu}, and \hyperlink{mailto:konstan@umn.edu}{\color{black}konstan@umn.edu}. \\
 This paper was originally circulated as ``The Economics of Recommender Systems: Evidence from a Field Experiment on MovieLens.'' 
 We thank 
 Heski Bar-Isaac, 
 Yeon-Koo Che, 
 Brett Gordon, 
 Ilya Morozov,  
 Shan Sikdar, 
 Andrey Simonov, 
 Greg Taylor, 
 Hal Varian, 
 and 
 Julian Wright 
 for their thoughtful feedback, 
 as well as audiences at the 
 AFE Conference (Chicago), 
 AIM Conference (USC Marshall), 
 ACM EC (KCL), 
 CESifo Economics of Digitization, 
 Chicago Booth, 
 Columbia, 
 Compass Lexecon, 
 EARIE (LUISS), 
 National University of Singapore, 
 NBER Workshop of Digital Economics, 
 Northwestern Institute on Complex Systems,
 Northwestern Kellogg,
 Psychology of Technology Conference (BU),
 PUC Chile, 
 UK Workshop on Digital Economics, 
 and 
 the Virtual Quant Marketing Seminar.
 Daniel Kluver and Joseph Konstan are members of the GroupLens lab, which runs and maintains the MovieLens platform; Ruoyan Kong was a member of the GroupLens lab. 
 This study was approved under both the Columbia IRB under protocol number AAAT2659 and the University of Minnesota IRB under protocol number STUDY00010960. 
 This work was partially supported by NSF grant CNS-2016397. \\
	\emph{First posted draft}: 25 November 2022. 
 \emph{This draft}: 
 \DDMonthYYYY\today.
}
\vspace*{.5em}

\setcounter{footnote}{0} \renewcommand{\thefootnote}{\arabic{footnote}}

\begin{center} \textbf{\large Abstract} \end{center}
\vspace*{1em}
\noindent\makebox[\textwidth][c]{
 \begin{minipage}{.85\textwidth}
 \noindent
 We conduct a field experiment on a movie-recommendation platform to investigate whether and how online recommendations influence consumption choices. Using a within-subjects design, our experiment measures the causal effect of recommendations on consumption and decomposes the relative importance of two economic mechanisms: expanding consumers' consideration sets and providing information about their idiosyncratic match value. We find that the informational component exerts a stronger influence -- recommendations shape consumer beliefs, which in turn drive consumption, particularly among less experienced consumers. Our findings and experimental design provide valuable insights for the economic evaluation and optimisation of online recommendation systems.
 \\[.5em]
 \end{minipage}
}
\newpage

\setcounter{page}{1}

\section{Introduction}
\label{section:introduction}
Recommendation systems (RS) are nearly ubiquitous in the digital economy.
These are systems that combine data from multiple consumers and sources to produce often personalised information in the form of consumer-specific recommendations \citep{ResnickVarian1997ACM}.
They have a wide set of applications: from e-commerce \citep{DinersteinEinavLevinSundaresan2018AER} and labour markets \citep{Horton2017JLE}, to curation feeds on social media platforms \citep{Levy2021AER}, to cultural goods on media streaming platforms \citep{HoltzCarteretteChandarNazariCramerAral2020WP}, and to the articles served on news platforms \citep{ChiouTucker2017JEMS,ClaussenPeukertSen2023MS}.
While these systems have been shown to increase consumption and engagement across these different contexts, their purportedly large influence has spawned a debate about their welfare implications as well as precisely what these systems ought to optimise for in order to best assist consumers \citep{McNeeRiedlKonstan2006, KleinbergMullainathanRaghavan2023MS}. 
The key question underlying this debate is \emph{how} they influence consumption choices.

In this paper, we discuss a field experiment on a movie recommendation platform, MovieLens, that measures the causal effect of online recommendations on consumption and decomposes the economic mechanisms that drive their influence. 
As recommender systems are typically deployed in environments with large choice sets, one prominent mechanism is that they can make consumers \emph{consider} goods, including goods that they did not know about. 
Furthermore, as these systems are prevalent in markets with experience goods, they can provide \emph{information} on goods' idiosyncratic match value. 
We use a within-subjects design that allows us to separately measure the effect of these mechanisms on consumption. 
We find that, while the consideration channel plays a meaningful role in increasing consumption, the informational channel plays a more substantial and important role in driving the effects of recommendation.

Our experimental design addresses two main challenges in measuring the causal effects of recommendations and disentangling these two channels, while being general enough to be implemented on any online platform with a recommendation system.

The first challenge is identifying the causal impact of recommendations on consumption, which is complicated by the fact that recommendations are targeted to consumers and so the experimental design needs to account for this selection while still providing high quality recommendations. 
A typical experimental design within this literature compares consumption between a group of consumers who receive recommendations from the recommender system algorithm and another who receive recommendations from reasonable alternative algorithms \citep{HoltzCarteretteChandarNazariCramerAral2020WP, KorganbekovaZuber2023WP}. 
While this design works well for assessing the effect of the recommendation system, it leads to a systematic bias when measuring the effect of recommendation as an RS should, by design, result in a treatment group with systematically higher idiosyncratic quality. 
Our experimental design exploits a unique aspect of our data to isolate the role of recommendation as we observe the intermediate outputs of the RS algorithm that provide us with an estimate of consumer-specific quality for each good. 
This motivates a within-subjects design that uses these estimates to generate for each consumer a \emph{control} group of goods, deliberately excluded from recommendations, and a \emph{recommendation} group of goods selected for recommendation, both of which have similar ex-ante idiosyncratic quality according to the RS estimates.

The second challenge is that consumer beliefs are typically unobserved and that recommendations simultaneously make consumers consider goods and provide information about good quality. 
We address the measurement issue with a belief elicitation survey that captures consumers' beliefs about the quality of unconsumed goods and how certain they are about these assessments. 
In this survey consumers only see the name of the good and its movie poster -- they cannot see more details about the good or the platform's predicted rating. A byproduct of this elicitation is that it makes consumers consider the good, without providing the informational content of recommendation.\footnote{
    As with much of the literature on consideration sets -- e.g., \citet{MasatliogluNakajimaOzbay2012AER} and \citet{ManziniMariotti2014Ecta} -- we do not distinguish between unawareness of some alternatives or working memory costs or limitations that require consumers to focus on a particular subset of the available goods.
} 
We exploit the variation induced from this survey to include a third \emph{consideration-only} group of goods for each consumer, which only show up in the belief elicitation survey but not in the recommendations. By carefully selecting the set of goods to elicit beliefs about and comparing consumption frequencies across groups, we measure the causal increase in consumption due to recommendation and quantify its informational gains.\footnote{
    Throughout the paper we are agnostic to whether the informational gains come from direct inferences that consumers make as a result of being recommended an item and indirect information accrued by the fact that recommendations reduce information acquisition costs.
}

Our experiment is conducted on a movie-recommendation platform, MovieLens, which has existed since 1997. 
MovieLens is noncommercial and devoted to producing helpful recommendations (it does not host movies) and features open-sourced data and algorithm implementation. Its data constitute a central benchmark in the recommender system community for the development and evaluation of new recommender system algorithms, having been used in thousands of papers.\footnote{
 Two examples in economics are \citet{ChenHarperKonstanLi2010AER} and \citet{Rossi2021}. 
 The vast majority of papers using MovieLens data rely on the ratings dataset to evaluate the performance of new recommendation system algorithms -- see \citet{HarperKonstan2015ACMTIIS} for an overview.
}
Specifically, the platform uses past ratings paired with a collaborative filtering algorithm \citep{HarperKonstan2015ACMTIIS} to produce \emph{user-specific} predicted ratings for unrated movies. These predictions are displayed to users in the first row of the platform homepage and are used to tailor the platform's \emph{user-specific} recommendations (``top picks'').
As such, the recommendations are personalised and considered high-quality. We maintain the high quality of recommendations during our intervention by including only the top 750 goods for each consumer
as determined by the RS estimate.\footnote{
    The focus is thus on estimating the effect of recommendations on consumption where the set of recommendations comprises ``high quality'' recommendations coming from the platform's recommendation algorithm and not randomly selected movies, which would not contain an informational component.
}

Our first main finding is that both consideration and recommendation induce a significant increase in consumption. 
Under our preferred specification, we find that consideration alone leads to a 0.2 percentage point (p.p.) increase in consumption relative to a baseline of 1.0 p.p. consumption of goods in the control group. 
In contrast, recommendation leads to a 1.3 p.p. increase in consumption relative to the consideration group, indicating that recommendation nearly doubles the probability of consumption relative to consideration alone.

Our second set of findings indicates that the larger increase in consumption from recommendations is primarily driven by their informational role. 
We show that recommendations causally influence beliefs, and that these, in turn, causally impact consumption decisions.

In order to causally identify how changes in beliefs drive consumption, we leverage the fact that our experimental intervention induces exogenous variation in recommendations. 
Realising that such variation constitutes randomised information provision \citep{HaalandRothWohlfart2023JEL} is the basis of our identification strategy, relying on an instrumental-variable approach. 
We find that a one-point increase in expected match value and a decrease in uncertainty (both on a 1-5 scale) lead to an 8.2 and 12.6 percentage point increase in consumption, respectively.

Having shown that beliefs drive consumption, we then turn to examining how recommendations affect beliefs.
We find that recommendations reduce uncertainty by 0.063 and shift expected match value assessments closer to the platform's predicted rating by 0.015. 
Paired with the earlier estimates on the relationship between beliefs and consumption, these changes in beliefs account for much of the observed increase in consumption. 
Additionally, less experienced consumers -- those with shorter consumption histories -- are more uncertain in their assessments and experience a larger causal increase in consumption from recommendations. 
This demonstrates that the informational role of recommendations operates as a function of the experience the consumer has, with newer consumers benefiting the most.

These results have important implications for understanding the welfare consequences of online recommendations.
Specifically, they imply that recommendations do more than manipulate consideration sets: they provide consumer-specific match value information in markets with experience goods. 
This suggests that, so long as the information provision from the recommendation system is unbiased,\footnote{
    As is the case in our context since the platform is noncommercial.
} 
recommendations improve consumer welfare largely through making them more informed about their match values, and not through expanding consideration sets. 
This suggests that the increasing trend of online platforms to introduce bias into their recommendation systems by steering consumers to more profitable goods can mainly hurt consumer welfare to the extent that it erodes information provision \citep{ArmstrongVickersZhou2009RAND, HagiuJullien2011RAND, CalvanoCalzolariDenicoloPastorello2022WP}.

Understanding the economic mechanisms through which recommendations influence choice is crucial not only for assessing their welfare implications but also for guiding the design of recommendation systems aimed at maximising consumer welfare. 
While the idea that recommendations primarily function as information provision is economically intuitive, it contrasts with a dominant focus in the literature on expanding consumers' consideration sets as the hallmark of ``good'' recommendations \citep{CastellsHurleyVargas2015, KaminskasBridge2016TiiS, Steck2018}. 
Our findings suggest that recommendations designed to provide match value information about considered goods, as emphasised by the serendipity evaluation criterion \citep{KotkovWangVeijalainen2016KBS}, are more effective at driving consumption and delivering what are often perceived as ``better'' recommendations. Moreover, our findings validate the importance of collecting economically motivated belief data, beyond traditional consumption data, as a tool for online platforms to directly measure and optimise for the informativeness of recommendations.\footnote{
    In a companion paper, we provide a scalable procedure for collecting such data, enabling their integration into platform recommender systems \citep{aridor2024movielens}.
}

\subsection{Related Literature}
\label{section:introduction:related-literature}

This paper contributes to a burgeoning literature on recommender systems and the impact of recommendations.

\subsubsection{The Impact of Recommendation on Consumption} Recent literature has examined whether recommendation systems impact consumption patterns.
\citet{SenecalNantel2004}, \citet{DasDatarGargRajaram2007}, \citet{FreyneJacoviGuyGeyer2009}, \citet{ZhouKhemmaratGao2010}, \citet{ClaussenPeukertSen2023MS}, \citet{HoltzCarteretteChandarNazariCramerAral2020WP}, and \citet{DonnellyKanodiaMorozov2024MS}
show that recommendation systems increased consumption in hypothetical choices in a lab experiment, Google News, a social network, a news website, YouTube, Spotify, and Wayfair respectively. 
These papers collectively provide evidence that personalised recommendations, relative to a non-personalised benchmark, meaningfully increase consumption. 
There has subsequently been a significant amount of work measuring the aggregate effects of recommendation systems on the types of goods that get consumed \citep{FlederHosanagar2009MS, NguyenHuiHarperTerveenKonstan2014, BrynjolfssonHuSimester2011MS,HosanagarFlederLeeBuja2013MS, LeeHosanagar2019ISR, HoltzCarteretteChandarNazariCramerAral2020WP, KorganbekovaZuber2023WP} with relevant implications for market competition and product variety, as well as the emergence of filter bubbles and echo chambers.

Our paper differs from both of these streams of literature by identifying the effect of \emph{recommendations} produced by a recommender system on consumption via good-level randomisation and by examining the mechanisms through which recommendations operate. 
Existing work compares consumption between personalised and non-personalised (i.e., popularity-based) recommendations. 
This experimental design is natural for assessing the impact of the recommendation algorithm on behaviour, but \emph{not} for isolating the role of or discerning the mechanisms for what the recommendation itself is doing.\footnote{
    This is since a well-implemented recommender system should, by design, result in a treatment group with systematically higher idiosyncratic quality relative to the control group, which biases measurement of the effect of recommendation itself.
} 
Our experimental design instead focuses on isolating the role of the recommendation itself by having a control and a consideration-only group of goods \emph{that would have been recommended}, which allows us both to discern the causal effect of recommendation and to disentangle its mechanisms. 
Indeed, the beliefs data we collect allows us to explain why recommendations may influence consumption -- crucial for understanding its welfare effects and for guiding design. Finally, this allows us to explore heterogeneity in the effects of recommendation across consumers.

Two related papers to ours within this stream of literature are \citet{kawaguchi2021designing} and \citet{ChenChanZhangLiuWu2023WP}. 
\citet{kawaguchi2021designing} decompose recommendation effects into attention and utility in a vending machine context. 
Our approach differs in two key ways: we directly measure information provision by eliciting consumer beliefs, and we focus on online platforms, where mechanisms differ significantly from the time-pressure dynamics emphasised in vending machines.
A more recent paper, \citet{ChenChanZhangLiuWu2023WP}, studies the effects of recommendations on consumer search patterns using a large-scale field experiment that changes the quality of recommendations and measures search intensity. 
In contrast, our work directly measures the informational gains from recommendations and separates these gains from their effects on consideration, providing a clearer understanding of their causal impact by controlling for selection rather than relying on exogenous variation in quality.

\subsubsection{Recommender System Evaluation} 
Our work contributes to the computer science literature on evaluating recommendation quality and determining which recommendations to present to consumers. Early studies in this field recognised that goods with the highest predicted ratings are not always the most useful recommendations \citep{McNeeRiedlKonstan2006}. Since then, various metrics have been proposed to assess recommendation quality by combining predicted consumer ratings with their past consumption history. These metrics can be categorised into two groups: those that define good recommendations as ones that expand consumers' consideration sets, such as coverage and novelty \citep{KaminskasBridge2016TiiS, CastellsHurleyVargas2015}, and those that prioritise providing ``unexpected and useful'' information on match value, such as serendipity \citep{KotkovWangVeijalainen2016KBS}. The rationale behind these metrics is that they implicitly incorporate theories about how recommendations influence consumer decision-making processes. Our work contributes to this literature by formalising how recommendations influence decision-making through an economic lens and by using a field experiment to quantify the relative importance of these mechanisms, ultimately guiding the choice between different evaluation metrics.

\subsubsection{Other Marketing Tools} 
There is a connection between the economic mechanisms that drive the effect of online recommendations and other marketing tools, such as advertising and consumer reviews. 
We defer the full discussion of the relationship between these to \hyref{online_appendix:additional_related_work}, but broadly argue that, despite sharing common mechanisms, recommendations are generated via an economically distinct process which leads to a relative difference in the importance of different mechanisms.

\section{Framework and Hypotheses}
\label{section:framework}

\subsection{Framework}
\label{section:framework:framework}
This section outlines a simple theoretical framework for modelling consumption decisions in the presence of recommendations. 
Throughout, we focus on the case of experience goods since it represents a common setting in which recommendation systems are deployed and is a good match for the environment of our experimental intervention. 
Our experiment will allow us to test both assumptions and implications of the model.

\subsubsection{Consumers}
We consider a consumer $i$ who, in each period $t=0,1,...$, makes a choice $c_t$, consisting of a good chosen from a finite set of goods $X$ or an outside option. 
We denote consumer $i$'s consumption history at time $t$ by $C_{i,t}:=\{c_{i,\ell} \in X, \ell \leq t\}$, and assume that consumers choose each good at most once.

\subsubsection{Match Value and Consumers' Beliefs}
Different consumers may value the same good differently. 
Each good $x$ has an idiosyncratic match value $v_{i,x}$ for each consumer $i$, reflecting their subjective assessment of its idiosyncratic value.
Prior to consuming a good, consumers are uncertain about its match value. 
We model this by assuming ${(v_{i,x})}_x$ are jointly normally distributed, with mean ${(v_{i,x}^b)}_x$ and covariance $\Sigma_i=[\sigma_{i,x,x'}^b]$, where $\sigma_{i,x,x'}^b\geq 0$. 
Upon consuming a good, consumer $i$ fully resolves uncertainty regarding its value, $v_{i,x}$.

\subsubsection{Consideration and Exposure}
Consumers may not consider all goods at all times.
We denote consumer $i$'s consideration set at time $t$ as $\Gamma_{i,t}$, which -- following, e.g., \citet{Goeree2008Ecta} and \citet{MasatliogluNakajimaOzbay2012AER} -- we take to represent the subset of goods the consumer is considering.
This can be due to limitations in working memory or simply being unaware of all options; similarly to existing literature,\footnote{
    A recent literature focuses on models of consideration set formation which give rise to probabilistic consideration of goods, e.g., \citet{CaplinDeanLeahy2019REStud}, \citet{CattaneoMaMasatliogluSuleymanov2020JPE}, and \citet{AbaluckAdams2021QJE}.
    We do not consider specific mechanisms through which consideration sets are refined, such as through shortlisting \citep{ManziniMariotti2014Ecta}, or costly search \citep{HonkaHortacsuVitorino2017RAND}.
} we do not take a stance on how the consumer's consideration set is formed. 
We focus instead on how recommendations may affect it.
 
Platforms manipulate consideration sets by prominently displaying specific goods and forcing consumers to pay attention to them. 
We consider such manipulations as $e_{i,x,t}=1$ to denote the situation in which the recommender system forces exposure of consumer $i$ to good $x$ at time $t$. 
Forced exposure implies consideration ($e_{i,x,t}=1 \implies x \in \Gamma_{i,t}$), but not the converse: goods that consumers were forcibly exposed to are a subset of the goods in their consideration set. 
This mechanism is thus similar to what has been considered in examining the effects of advertising \citep[see, e.g.,][]{Goeree2008Ecta}.

\subsubsection{Learning and Recommendation}
Recommender systems typically affect consumers' information beyond the formation of their consideration set. 
Since recommendations correspond to prominently featuring a subset of goods as being of high predicted match value (``top picks for you''), being recommended can itself signal a higher (match) value.
Indeed these recommendations are driven by the recommender system's personalised estimate of the value of each good $x$ for consumer $i$, denoted $v_{i,x}^p$, which is also often displayed.
We will term such signal the \emph{platform's predicted value}.
We model this direct information provision and denote a recommendation for good $x$ to consumer $i$ at time $t$ by $r_{i,x,t} = 1$ and assume that the platform discloses its consumer-idiosyncratic predicted value $v_{i,x}^p$. 
For simplicity, $v_{i,x}^p$ is modelled as a Gaussian signal centred on $v_{i,x}$.

In addition to providing direct information about the value of a good, recommendations can have a further informational effect. 
Specifically, due to prominent positioning, recommendations reduce information acquisition costs by making it easier to learn about the good's attributes. We consider both of these components as the informational aspect of recommendation and do not decompose their relative importance.

\subsubsection{Choice}
Consumer $i$ evaluates match value according to an increasing and concave utility function $u_i:\mathbb R\to \mathbb R$, capturing the consumer's attitudes toward uncertainty.
We assume the utility associated with the outside option available at time $t$, $u_{i,o}$, is distributed according to some distribution $F$, independently across match values and across periods, and known to the consumer before making their choice at time $t$.
Then, given their information at time $t$, consumer $i$ at time $t$ chooses to maximise current expected utility, i.e., $\displaystyle c_{i,t}\in \argmax_{y \in \Gamma_{i,t}\cup \{o\}} \mathbb E_{i,t}[u_i(v_{i,y})]$.

\subsection{Hypotheses}
\label{section:framework:hypotheses}
Recommendations affect consumption through multiple informational channels. 
One important issue is that, by recommending good $x$ to consumer $i$, the platform is also exposing the consumer to that good and therefore forcing consideration ($r_{i,x,t}=1\implies e_{i,x,t}=1 \implies x\in\Gamma_{i,t}$). 

Our first hypothesis is that forcing consideration has on its own a positive impact on consumption, but recommendations affect it further:
\begin{hypothesis}
 \label{hypothesis:consumption-on-rec}
 Forcing consideration of a good increases its consumption; recommending it increases consumption further.
\end{hypothesis}

Our remaining hypotheses consider whether recommendations' effect on consumption can be partly explained by their role in directly providing information about the goods' match value.
Underlying our theoretical framework is the assumption that recommendation affects consumption by affecting consumers' beliefs, which in turn explain consumption patterns.
We decompose our analysis in two steps.

First, we test whether beliefs are sensibly related to consumption. 
In particular, one would expect that consumers are more likely to choose goods they believe have a higher expected value, $v_{i,x,t}^b:=\mathbb E_{i,t}[v_{i,x}]$, which in our framework corresponds to utility being increasing in value.
On the other hand, if consumers are uncertainty-averse (concave utility), then, all else equal, they are more likely to choose goods about whose match value they are less uncertain.
We then have:
\begin{hypothesis}
 \label{hypothesis:consumption-on-beliefs}
 Goods with higher expected match value and lower uncertainty are more likely to be consumed.
\end{hypothesis}

Then, we test the effect of recommendations on consumers' beliefs.
In our theoretical framework, we assumed that the recommendation of good $x$ to consumer $i$ at a given time provides a noisy signal of the true match value, that is, $v_{i,x}^p=v_{i,x}+\sigma_{i,x}^p\,\epsilon_{i,x}^p$, where $\epsilon_{i,x}^p\sim N(0,1)$.\footnote{
    The recommendation system provides recommendations that are among the platform's best guesses for the consumer's preferred good and as such we expect updating to mostly be in the positive direction. 
    However, this is not always the case and, for generality, we allow updating to go in either direction.
} 
This not only leads to the consumer being more certain about their valuation of good $x$ but also drives their expected value toward the signal. 
We then posit the following:
\begin{hypothesis}
    \label{hypothesis:beliefs-on-rec}
    Recommending a good (i) makes consumers less uncertain and (ii) drives their beliefs towards the platform's predicted match value.
\end{hypothesis}

\section{Experimental Design}
\label{section:design}
In order to study if and how recommendations impact consumption we conduct an experimental intervention on a movie recommendation online platform, MovieLens.
Our intervention has two main features: 
(i) we generate random variation in recommendations to study their causal effect on consumption, and 
(ii) we elicit belief data about good match value prior to consumption to examine the informational mechanisms through which recommendation acts.
In this section, we provide background information on the platform and describe our experimental procedures.

\subsection{Background on the Recommendation Platform}
\label{section:design:background-on-platform}
MovieLens is a movie-recommendation platform created in 1997. 
It is used by consumers to obtain information about movies as well as personalised movie recommendations based on their ratings. 
The platform has been widely used, and its movie ratings data are a central benchmark in the recommender system community for the evaluation of new recommender system algorithms.\footnote{
    For instance, the search expression ``MovieLens dataset'' or ``MovieLens data'' returns over 9,000 entries on Google Scholar, whereas ``Netflix dataset'' or ``Netflix data'' -- which includes both proprietary data and the well-known public access Netflix Prize competition \citep{BennetLanning2007} -- returns less than half the number of entries.
}

The platform's home page displays movies organised by categories in rows, with the very first one showing eight ``top picks'', the platform's top recommended movies for the user. 
Movies are set in a grid fashion, with their poster, title, and the platform-predicted rating for the user.\footnote{
    Screenshots of the platform's interface are included in the \hyref{online-appendix:screenshots}.
} 
The platform-predicted ratings are personalised to each user and obtained through a sophisticated collaborative filtering algorithm combining the user's and others' past ratings, reviews, and other metadata \citep{HarperKonstan2015ACMTIIS}. 
When hovering over a movie title, users see its genres, their platform-predicted rating, and the average and number of community ratings. 
Subsequent rows correspond to recent releases, and other categories of potential interest (e.g., ``favorites from the past year'' or ``new additions'').

Consumers use the platform to find movies to watch and to rate the movies after watching. 
Clicking on a movie's page provides access to detailed information about the movie, including its trailer, synopsis, cast, associated tags \citep{VigSenRiedl2012ACM}, and similar movies.
It does not provide consumption opportunities, as it does not host movies to stream nor does it direct users to other platforms.
This allows us to study consideration effects of recommendations separately from search costs, since recommendations do not affect potential frictions in finding or accessing a particular target consumption good. 
Furthermore, the platform is free to use and noncommercial, and it is incentive-compatible for users to truthfully report their ratings, as truthful information improves the platform's recommendation quality.

The recommendation system used by the platform is of high quality and ideal from both a user's and a researcher's perspective for several reasons. 
First, its open-source ratings data are extensively used to develop and evaluate high-quality recommendation algorithms, which ultimately support the quality of the recommendation algorithms MovieLens deploys. 
Second, the set of algorithms used is transparent, as these are open source and constitute canonical implementations of widely used item-item or singular-value-decomposition collaborative filtering algorithms \citep[see][]{EkstrandLudwigKolbRiedl2011}. 
Finally, MovieLens operates as a noncommercial platform, aligning its recommendation focus with user satisfaction rather than platform profitability. 
This user-centric approach alleviates concerns about potential biases in user perception of recommendations.

\subsection{Experimental Intervention}
\label{section:design:experimental-intervention}
The platform provides a natural setting to test our hypotheses.
We take ratings as a measure of realised good match value and the platform's predicted rating as a noisy signal; we will use rating and (match) value interchangeably.
We take the platform's `top picks' category as a recommendation of the specific goods listed there.

To understand recommendations' informational impact, we regularly survey consumers' beliefs. 
We identify the effect of recommendations by inducing exogenous randomness and comparing outcomes for recommended goods to those which \emph{would} otherwise be as likely to be recommended.

At the start of the experiment, consumer $i$'s 750 goods (not previously consumed) with the highest platform-predicted value are split into three sets: a \emph{control set}, a \emph{consideration-only set}, and a \emph{recommendation set}. 
We only elicit beliefs about match value for goods in the consideration-only and the recommendation sets, and we restrict recommendations to goods in the latter. 
We control for consumer's idiosyncratic tastes by employing stratified block randomisation: we randomly assign the $n$-th, $(n+1)$-th, and $(n+2)$-th goods with the highest platform-predicted value to each of the three sets.

Every day that the consumer enters the platform, we elicit their beliefs about the match value of 10 goods not previously consumed.\footnote{
    The survey can be deferred to the following visit to the platform. 
    The set of platform recommendations remains the same until the survey is completed. 
} 
After completing the survey, the consumer is taken to the platform's home page, in which they are presented with the set of eight recommended goods (their ``top picks'').
The following procedure summarises how we choose the goods for belief elicitation and recommendation.
\begin{proc}[Belief Elicitation and Recommendation]
    \label{procedure:belief-elicitation-rec}~
    \begin{enumerate}[label=Step \arabic*., itemsep=0em]
        \item 
        Elicit good consumption occurring since the previous visit and remove the consumed goods from the control, consideration-only, and recommendation sets. 
        \item 
        For both the consideration-only and recommendation sets, 
        select uniformly at random two out of the top eight goods, resorted by platform-predicted value plus i.i.d. Gaussian noise. 
        \item 
        Elicit beliefs about the four goods selected in Step 2 in the current period, the four selected in the previous period, and two more selected uniformly at random from the consideration-only set.
        \item
        Recommend the top 8 goods in the recommendation set as per the sorting generated in Step 2.
    \end{enumerate}
\end{proc}

The algorithm underlying the predicted rating remained the same. 
Recommendations are therefore credible, as these refer to goods with higher predicted value on average (Steps 2 and 5). 
As recommended goods are of high predicted value, our sampling procedure elicits beliefs on goods of similar predicted value in the consideration-only set (Step 3). 
This, together with having beliefs about goods being elicited in two subsequent periods (Step 4), enables us to identify how recommendations impact beliefs. 
By randomly selecting two goods from the consideration-only set to elicit beliefs on (Step 4), we learn consumers' beliefs about goods for a broader part of the domain.

Finally, we note that this procedure also allows us to isolate the effects of forced consideration. 
Specifically, we leverage the unavoidable fact that eliciting beliefs about a good's match value requires exposing consumers to that good to identify the effect of consideration by comparing outcomes for goods in the consideration-only set and those in the control set.\footnote{
    We highlight that, via our elicitation survey, our consideration manipulation unambiguously makes the consumer aware of and actively consider the good in question, whereas on the platform consumers may overlook or ignore some recommendations and thus not actively consider them  \citep{ZhaoChangHarperKonstan2016}. 
    Hence, our estimates likely represent an upper bound on the effect of consideration.
}

\subsection{Measurements}
\label{section:design:measurements}
In this section, we provide details on the belief data and consumption measurement.

\subsubsection{Beliefs about Match Value}
\label{section:design:measurements:beliefs}
We elicit consumers' beliefs about match value through a survey they are asked to fill out whenever they return to the platform. 
In the survey, we elicit beliefs about the match value of 10 movies, selected according to the procedure described in \hyref{section:design:experimental-intervention}[Section].

For each of the 10 selected movies, the consumer was asked whether or not they had watched it. 
If they had watched it, we asked for their rating -- corresponding to $v_{i,x}$, the idiosyncratic realised match value -- and an approximate date of when they watched it. 
If they had not watched it, we elicited both their expected rating and a measure of uncertainty about their reported expectation.

The expected rating serves as a measure of expected match value, corresponding to $v_{i,x,t}^b$, and is measured on the same scale as the one used to rate movies in the platform (a 10-point Likert scale, ranging from 0.5 to 5). 
Our measure of uncertainty $\sigma_{i,x,t}^b$ is proxied by asking how sure participants are of their reported expected rating on a 5-point Likert scale, ranging 1-5, where higher values are associated with higher uncertainty about the expected rating.
As in our theoretical framework, we assume that, upon consumption, consumers perfectly learn the match value of the good, $v_{i,x,t}^b=v_{i,x}$, and there is no uncertainty, $\sigma_{i,x,t}^b=0$. 
We then take the reported ratings as part of our belief data.

The survey interface matches closely the platform's interface for ratings, except in omitting the predicted rating and not displaying information other than the name and movie poster.\footnote{
    See \hyref{online-appendix:screenshots} for interface screenshots.
} 
Unlike on the homepage, it is not possible for the consumer to hover over or click into the details page to acquire additional information on the good.

We assume belief data were truthfully reported. 
While the survey was not incentivised, consumers do not have an incentive to misreport: the platform is noncommercial and helping the platform improve its recommendation system is in the consumers' own interest. 
The analysis of the belief data provides assurance that the belief data are internally consistent and consistent with the consumers' behaviour on the platform.
In \ref{appendix:beliefs-data}, we show that \emph{(i)} consumers' initial estimate of expected match value (i.e., the first elicitation before any experimental intervention) on average equals the realised match value rating they provide after consumption, and \emph{(ii)} the distance between consumers' expected and realised match value is increasing in our uncertainty measure.
We also show that uncertainty about expected match value is decreasing \emph{(a)} in consumers' experience, measured by the number of movies rated at the outset of the intervention period, \emph{(b)} in the movie's popularity, proxied by the number of community ratings, and \emph{(c)} on whether the movie is a sequel or part of a franchise; see \hyref{table:reg-beliefs-on-popularity}[Table].

\subsubsection{Good Consumption}
\label{section:design:measurements:consumption}
The platform does not offer consumption opportunities and therefore there is no direct observation of consumption.
The most natural proxy for consumption is to rely on ratings that consumers input on the platform to determine what is consumed and when. 
Since we are concerned with treatment effects on consumption during our intervention, we exclude ratings referring to consumption that occurred prior to the start of the intervention, as per the report of approximate consumption date in the survey; our results are nevertheless robust to also including consumption occurring prior to the intervention.

\subsection{Recruitment and Study Implementation}
\label{section:design:other-details:recruitment}

In our intervention, we targeted a random sample from a subset of the platform's users.\footnote{
    The use of or access to the platform is prohibited to individuals under the age of 18, as per the platform's terms of service.
}
We sought to mitigate the heterogeneity of treatment effects across consumers arising from differences in the quality of the recommendations.
As such, we recruited users with a minimum level of engagement, so that the recommender system algorithm used by the platform is able to produce valuable recommendations, while avoiding power users who would constitute extreme outliers. 
The specific criteria, laid out in \hyref{appendix:recruitment}, were chosen with the platform's experts.

The intervention lasted from 29 March 2021 to 31 October 2021. 
It had a phased rollout, including a randomly selected portion of consumers from 29 March 2021 until 15 April 2021, followed by a full rollout to all eligible consumers.
The length of the study period was selected based on power calculations and considering the possibly slow rate of consumption of movies over time.
The intervention targeted 4,572 eligible individuals, of which 1,452 decided to enroll in the study; we conduct our analysis over these participants. 
During the intervention, on average consumers in our study visited the platform on 11.67 days, watched 24.02 movies, and completed 2.84 belief surveys.

\section{The Impact of Recommendations on Consumption}
\label{section:consumption-on-rec}

In this section, we test \hyref{hypothesis:consumption-on-rec}[Hypothesis]: whether consideration induces additional consumption and whether recommendation further influences consumption beyond consideration.
Our baseline specification is as follows:
\begin{align}
    \label{equation:direct-rec-on-consumption1}
    c_{i,x} = \beta_0 + \beta_1 e_{i,x} + \beta_2 r_{i,x} + \epsilon_{i,x} 
\end{align}
\noindent 
where $c_{i,x}=1$ if, during the intervention, consumer $i$ reported having consumed good $x$ (and zero if otherwise), $e_{i,x}=1$ if good $x$ is in consumer $i$'s consideration-only or recommendation sets and zero if it is in the control set, and $r_{i,x}=1$ if $x$ is in $i$'s recommendation set.

Recall that consideration occurs for consumer $i$ through belief elicitation of good $x$ only if $x$ is in the consideration-only or recommendation sets, and platform recommendation of good $x$ occurs only if $x$ is in the recommendation set.
Naturally, the consumer may have other recommendation sources and, even through the platform, may be exposed to goods in either of these sets.
Since we stratified our randomisation by consumer-specific tastes, exposure and recommendation to goods via other channels should be orthogonal to treatment assignment.

While this specification enables a clear and straightforward causal estimate of the impact of consideration and recommendation on consumption, it is potentially too conservative.
Specifically, it does not take into account the fact that some goods in the recommendation set are never explicitly recommended and some goods in the consideration-only set are never selected for belief elicitation. 
In order to obtain better estimates on the treatment effects, we consider two additional strategies.

First, we consider the same specification (\ref{equation:direct-rec-on-consumption1}), but we restrict the sample to goods in exposed randomisation blocks.
While we would like to compare the effect of actual  consideration and actual recommendation to the control group, realisations of consideration and recommendation are not independent of platform-predicted match value.\footnote{
    A design restriction, since recommendations need to remain useful and meaningful to consumers.
} In order to resolve this issue, we take not only goods that the consumers were actually exposed to through belief elicitation or recommendation but also goods in their randomisation blocks.
Recall that, at the outset of the intervention, goods with the $n$-th, ($n$+1)-th, or ($n$+2)-th highest platform-predicted match value for consumer $i$ are bundled into the same block and block-randomised into the control, consideration-only, and recommendation sets. 
Then, if good $x$ was recommended ($r_{i,x}=1$) to consumer $i$ or exposed to it through our experimental intervention ($e_{i,x}=1$), we include $x$ in the sample, as well as the other goods in the same block used for block-randomisation.
We then estimate the same specification (\ref{equation:direct-rec-on-consumption1}), but we obtain more precise estimates on the average treatment effect of consideration and recommendation on consumption.

Second, we estimate a variation of the same specification (\ref{equation:direct-rec-on-consumption1}), but we restrict the sample to goods that participants were actually exposed to and where they explicitly told us in the elicitation survey that they had not consumed them before.
As such, we only keep goods about which beliefs were elicited, that is, those in the consideration-only and recommendation sets. 
This, together with our block randomisation, guarantees that actual recommendation $r_{i,x}$ is orthogonal to the good's characteristics.

\setstretch{1}
\begin{figure}[tb]
    \centering
    \includegraphics[scale=0.6]{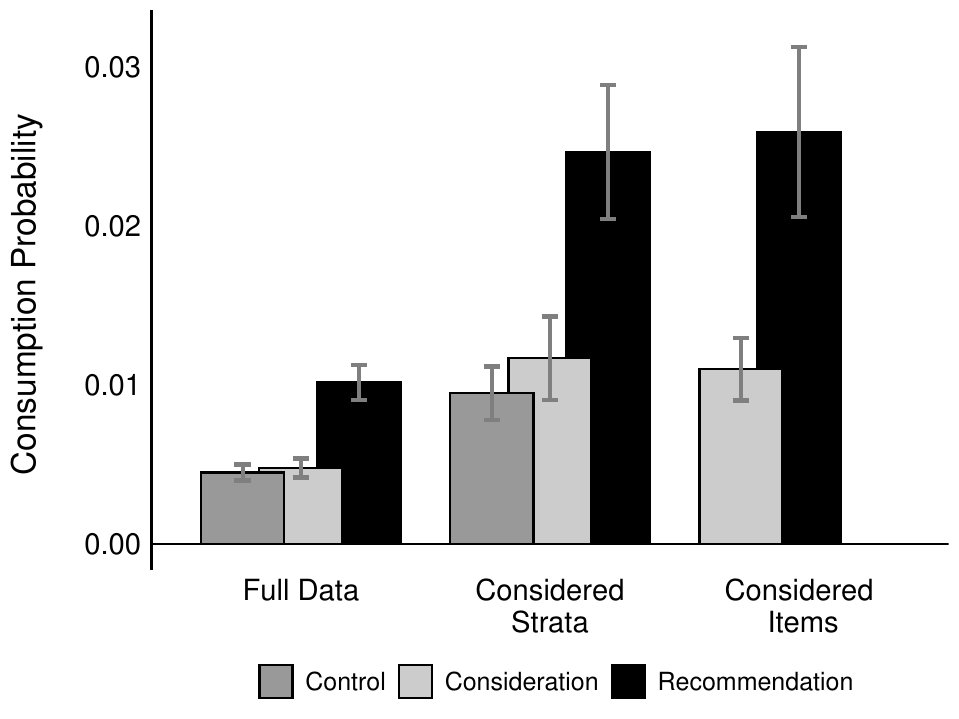}
    \begin{minipage}{1\linewidth}
        \vspace*{.5em}
        \caption{The Impact of Recommendation on Consumption}
        \label{figure:reg-consumption-on-rec}
        \small
        \setstretch{1} \emph{Notes}: 
        This figure tests whether consideration and recommendation impact consumption probability.
        The estimates correspond to those specified in \hyref{equation:direct-rec-on-consumption1}[Equation].
        Each component displays the baseline control and the estimated average treatment effect of consideration and recommendation on consumption for the different sample specifications. 
        Each observation corresponds to a pair (consumer $i$, good $x$). 
        ``Full Data'' includes all consumers $i$ and all goods $x$ in the consumer-specific control, consideration-only, and recommendation sets. 
        ``Considered Blocks'' includes the goods to which a consumer was exposed to through the belief elicitation survey, and all the goods in the same consumer-specific randomisation block. 
        ``Considered Items'' includes only goods to which a consumer was exposed through the belief elicitation survey; it necessarily excludes goods in the control set.
    \end{minipage}
\end{figure}
\setstretch{1.3}

We summarise the results across these different specifications in \hyref{figure:reg-consumption-on-rec}[Figure], where full data, considered (randomisation) blocks, and considered items correspond to the three estimation specifications discussed above.\footnote{
    The regression tables for the different specifications are presented in \hyref{table:reg-consumption-on-rec}[Table] in \hyref{online-appendix:additional-tables}[].
} Under our preferred specification -- restricting to considered blocks -- the baseline consumption probability of items in the control group was 1\% of the goods in the considered blocks. 
The effect of consideration alone results in a statistically significant 0.2 percentage point increase, which is a 20\% increase relative to the control, providing support for the first part of \hyref{hypothesis:consumption-on-rec}[Hypothesis]. 
Furthermore, the effect of recommendation leads to an \emph{additional} 1.3 percentage point increase and results in an aggregate 150\% increase relative to the control group, providing support for the second prediction of \hyref{hypothesis:consumption-on-rec}[Hypothesis]. 
Overall, even the magnitude of the impact of recommendation on consumption probability is fairly consistent across the different estimation strategies.

\subsection{Persuasion Rates}
\label{section:consumption-on-rec:persuasion}

Recommendations crucially depend on their credibility, on their ability to alter beliefs and, consequently, behaviour. 
One natural way of assessing the effect of recommendations on behaviour is by estimating their persuasion rates. 
Following \cite{DellaVignaGentzkow2010ARE}, we compute the persuasion rate of consideration and recommendations, respectively, $p_{\text{consideration}}$ and $p_{\text{recommendation}}$, as
\begin{align}
    p_T:=\frac{y_T-y_\text{control}}{e_T-e_\text{control}}\frac{1}{1-y_\text{control}},
\end{align}
where $y_T\in [0,1]$ denotes the share of goods in group $T \in \{\text{consideration}, \text{recommendation}\}$ which were ultimately consumed and $y_\text{control}$ that for the control group, while $e_\text{consideration}$ denotes the share of goods in the consideration treatment group that were forced consideration, and $e_\text{recommendation}$ the share of those in the recommendation group that were actually recommended.

We find that the persuasion rate of consideration is 10.34\%, while that of recommendations is 33.34\%.
Furthermore, if we consider the persuasion rate of consideration absent recommendation, that is, focusing on consideration only, the persuasion rate drastically drops to 0.73\%, indicating that the persuasion power is overwhelmingly through recommendations, and that consideration is not a significant force in driving consumption once one factors out actively recommended items.\footnote{
    Recall, from our earlier discussion, that we expect this to be an upper bound on the effect of consideration since our intervention forces consideration of each good.
} 
We also note that persuasion rates are relatively large compared to those found in other settings related to the persuasion of consumers \citep[see][]{DellaVignaGentzkow2010ARE}, in line with the predominance of the recommendation platform's non-strategic informational role.

\section{Recommendations, Beliefs, and Consumption}
\label{section:mechanisms}
The results from \hyref{section:consumption-on-rec}[Section] support \hyref{hypothesis:consumption-on-rec}[Hypothesis], indicating that, while consideration increases consumption, recommendation has a significantly larger effect. 
In this section, we examine whether the large effect of recommendation on consumption can be rationalised by an informational mechanism. 
Namely, we examine the causal impact of beliefs on consumption, and then that of recommendations on beliefs.

\subsection{Beliefs Explain Consumption}
\label{section:mechanisms:consumption-on-beliefs}

We start by evaluating the extent to which the belief data explains consumption behaviour. 
We test \hyref{hypothesis:consumption-on-beliefs}[Hypothesis], which -- in line with our theoretical framework -- suggests the likelihood of consumption is increasing in the expected match value and decreasing in reported uncertainty.

\setstretch{1}
\begin{table}[th!]\setstretch{1.1}
	\centering
	\begin{tabular}{l@{\extracolsep{4pt}}ccc@{}}
  \hline\hline
  \hspace*{9em} & \multicolumn{3}{c}{Consumption} \\
  \cline{2-4}
  \hspace*{9em} & (1) & (2) & (3) \\
  \hline
  Uncertainty & $-$0.139$^{***}$ &  & $-$0.126$^{**}$ \\ 
   & (0.054) &  & (0.054) \\ [.1em]
  Exp. Match Value &  & 0.061$^{***}$ & 0.082$^{***}$ \\ 
   &  & (0.019) & (0.029) \\ [.1em]
  Constant & 0.486$^{***}$ & $-$0.168$^{***}$ & 0.195 \\ 
   & (0.181) & (0.056) & (0.172) \\ [.5em]
  \hline 
  Weak Instruments: Uncertainty & 10.07 &  & 4.08 \\ ~
    & [0.002]  &  & [0.007]  \\  [.2em]
  Weak Instruments: Exp. Match Value &  & 55.40 & 20.20\\ ~
    &  & [0.000]  & [0.000]  \\  [.2em]
  Wu-Hausman & 15.71 & 8.39 & 11.29 \\ ~
    & [0.000]  & [0.004]  & [0.000]  \\  [.2em]
  Sargan &  &  & 0.18  \\ ~
    &  &  & [0.672]  \\  [.2em]
  Observations & 20,895 & 20,895 & 20,895 \\ 
  \hline\hline
  \multicolumn{4}{l}{\footnotesize Clustered standard errors at the subject level in parentheses.}\\
  \multicolumn{4}{l}{\footnotesize p-values in squared brackets.}\\
  \multicolumn{4}{l}{\footnotesize $^{*}$ \(p<0.1\), $^{**}$ \(p<0.05\), $^{***}$ \(p<0.01\)}
\end{tabular}
    \begin{minipage}{1\textwidth}
        \vspace*{.5em}
        \caption{Beliefs Explain Consumption (\hyref{hypothesis:consumption-on-beliefs}[Hypothesis])}
        \label{table:iv-consumption-on-beliefs}
        \setstretch{1} \emph{Notes}: 
        This table tests \hyref{hypothesis:consumption-on-beliefs}[Hypothesis] by estimating the causal effect of a good's expected match value and uncertainty on whether it is consumed, as per \hyref{equation:beliefs-on-consumption}[Equation]. 
        It considers only the last time beliefs about a specific good are elicited from a given consumer. 
        The instruments are as described in \hyref{section:mechanisms:consumption-on-beliefs}[Section]; Weak Instruments, Wu-Hausman, and Sargan, correspond to tests regarding weak instruments, endogeneity, and validity of overidentifying restrictions.
    \end{minipage}
\end{table}
\setstretch{1.3}

We evaluate the relationship through the following regression:
\begin{align}
    \label{equation:beliefs-on-consumption}
    &
    c_{i,x} = \beta_1 v_{i,x}^b + \beta_2 \sigma_{i,x}^b + \epsilon_{i,x}
\end{align}
\noindent
where $v_{i,x}^b$ and $\sigma_{i,x}^b$ denote the last elicitation of consumer $i$'s expected match value and uncertainty associated with good $x$.

In order to enable a causal interpretation of the relationship between beliefs and consumption, we rely on an instrumental variables approach. 
We instrument expected match value and uncertainty with two variables: 
(1) whether a belief elicitation occurred before or after a recommendation (\`{a} la \citet{HaalandRothWohlfart2023JEL}), exploiting the randomisation of which movies are recommended; and 
(2) the user's activity level, proxied by their study opt-in date, leveraging the phased rollout of the experiment and the orthogonality of platform usage to preferences. 
We include the interaction of both instruments when analysing both endogenous regressors.

Our results are displayed in \hyref{table:iv-consumption-on-beliefs}[Table]. 
Columns (1) and (2) present the results of using the information provision and enrollment instrumental variables separately for uncertainty and expected match value, respectively. 
Column (3) presents the results of estimating specification (\ref{equation:beliefs-on-consumption}) using both of the instrumental variables. 
The results are consistent across all specifications.
Specifically, the estimates suggest that a one-point increase in expected match value and decrease in uncertainty lead to, respectively, a 8.2 and a 12.6 percentage point increase in consumption probability, lending support for \hyref{hypothesis:consumption-on-beliefs}[Hypothesis]. 
These estimates are economically meaningful as they suggest that the large effect sizes observed in \hyref{section:consumption-on-rec}[Section] can be rationalised if recommendation induces changes to consumer beliefs.

We perform standard diagnostic tests to assess the validity of the instruments \citep{Wooldridge2010Econometric}. 
The F-statistics confirm that both IVs are strong, individually and jointly: 10.07 for uncertainty (Column 1), 55.40 for expected match value (Column 2), and 4.08 and 20.20, respectively, when combined with interactions (Column 3).
A Wu-Hausman test reveals significant correlations (p < 0.01) between the endogenous regressors (uncertainty, expected match value) and the error term in all specifications, indicating that OLS estimates are biased and IV estimation is necessary. 
Supporting this, \hyref{table:reg-consumption-on-beliefs}[Table] in the \ref{online-appendix:additional-tables} shows that OLS estimates significantly underestimate the effects of uncertainty and expected match value on consumption.
Finally, the Sargan test in Column 3 yields a nonsignificant result (p = 0.672), supporting the validity of the exclusion restriction.

\subsection{The Impact of Recommendations on Beliefs}
\label{section:mechanisms:beliefs-on-rec}

Now that we have established that beliefs causally determine consumption, 
we explore the extent to which recommendation impacts these beliefs as a possible explanation for the increase in consumption resulting from recommendation. 
The information conveyed by recommendations about potentially high match value goods can lead to belief updating and subsequently influence consumption decisions. 
This particular informational channel of recommendations constitutes the crux of \hyref{hypothesis:beliefs-on-rec}[Hypothesis].

We examine two different aspects.
We first test the first-order effect of whether recommendations provide information, that is, whether they decrease uncertainty. 
Second, we test if recommendations drive consumers' expected match value assessments closer to the platform's predicted match value, that is, whether recommendations decrease the consumer-platform expected match value gap.
We estimate the following specification:
\begin{align}
 \label{equation:rec-on-beliefs}
 y_{i,x} &= \beta_0 + \beta_1 r_{i,x} + \epsilon_{i,x}
\end{align}
where $r_{i,x}$ is an indicator that equals 1 if good $x$ was recommended to consumer $i$ and is otherwise 0, and $y_{i,x}=\Delta \sigma_{i,x}^b$ or $\Delta |v_{i,x}^p-v_{i,x}^b|$ which denote, respectively, the change in consumer $i$'s uncertainty about good $x$'s value and the change in the consumer-platform match value gap, i.e., the difference between consumer $i$'s expectation of good $x$'s match value, $v_{i,x}^b$, and the platform's predicted match value of good $x$ for consumer $i$, $v_{i,x}^p$.
The change is taken to be over the course of the experiment, considering the first and last beliefs reported for each consumer $i$ and good $x$.

\setstretch{1}
\begin{table}[t]\setstretch{1.1}
	\centering
	\begin{tabular}{l@{\extracolsep{4pt}}cc@{}}
  \hline\hline
  & $\Delta$ Uncertainty & $\Delta$ Consumer-Platform \\
  &                      & Exp. Match Value Gap \\
  \cline{2-3}
  \hspace*{9em} & (1) & (2) \\
  \hline
  Recommendation & $-$0.063$^{***}$ & $-$0.015$^{**}$ \\ 
   & (0.013) & (0.007) \\ [.1em]
  Constant & $-$0.055$^{***}$ & 0.004 \\ 
   & (0.008) & (0.004) \\ [.1em]
  \hline \\
  Observations & 21,283 & 21,283 \\ 
  R$^{2}$ & 0.001 & 0.0002 \\ 
  \hline\hline
  \multicolumn{3}{l}{\footnotesize Clustered standard errors at the subject level in parentheses.}\\
  \multicolumn{3}{l}{\footnotesize $^{*}$ \(p<0.1\), $^{**}$ \(p<0.05\), $^{***}$ \(p<0.01\)}
\end{tabular}
    \begin{minipage}{1\textwidth}
        \vspace*{.5em}
        \caption{The Impact of Recommendation on Beliefs (\hyref{hypothesis:beliefs-on-rec}[Hypothesis])}
        \label{table:reg-beliefs-on-rec}
        \setstretch{1} \emph{Notes}:
        Columns (1) and (2) test \hyref{hypothesis:beliefs-on-rec}[Hypothesis][(i)] and \hyref{hypothesis:beliefs-on-rec}[][(ii)], respectively, by estimating the effect of recommendations on the change in uncertainty $\Delta \sigma_{i,x}^b$ and on the change of the absolute difference between the consumer's expected value and the platform's predicted value, $\Delta |v_{i,x}^b-v_{i,x}^p|$, as per \hyref{equation:rec-on-beliefs}[Equation]. 
        The change is taken to be over the course of the experiment, considering the first and last beliefs reported for each consumer $i$ and good $x$.
    \end{minipage}
\end{table}
\setstretch{1.3}

Since goods selected for belief elicitation are of the same expected (high) match value, regardless of whether they are recommended or not, we are able to identify a causal effect of recommendations on consumer beliefs. 
Columns (1) and (2) of \hyref{table:reg-beliefs-on-rec}[Table] present our estimates for equation (\ref{equation:rec-on-beliefs}) with $y_{i,x}=\Delta \sigma_{i,x}^b$ and $y_{i,x}=\Delta |v_{i,x}^p-v_{i,x}^b|$, respectively. 
We find support for \hyref{hypothesis:beliefs-on-rec}[Hypothesis], with recommendations decreasing uncertainty for the recommended good and decreasing the match value difference between consumers' expectations and the platform's predictions.
Recommending an item significantly decreases uncertainty by .06 points, and closes the match value gap by .015 points. 
Put together with the effect sizes estimated in \hyref{section:mechanisms:consumption-on-beliefs}[Section], this shows that the increase in consumption due to recommendations can be rationalised through its effect on beliefs.

\section{Heterogeneous Effects and Robustness}

In this section, we explore heterogeneous effects of recommendation and conduct different robustness exercises. 

\subsection{Heterogeneous Effects}
\label{section:robustness:heterogeneity}

We first examine how the treatment effect depends on prior beliefs and previous experience.
If their informational role is a main mechanism through which recommendations operate on consumption, then recommendation should have a greater impact when consumers are less certain about their valuation of a good and when they are less experienced with the product space.

\subsubsection{Prior Beliefs}

\begin{figure}[t]
    \centering
    \begin{subfigure}{.45\textwidth}
        \includegraphics[width=1\linewidth]{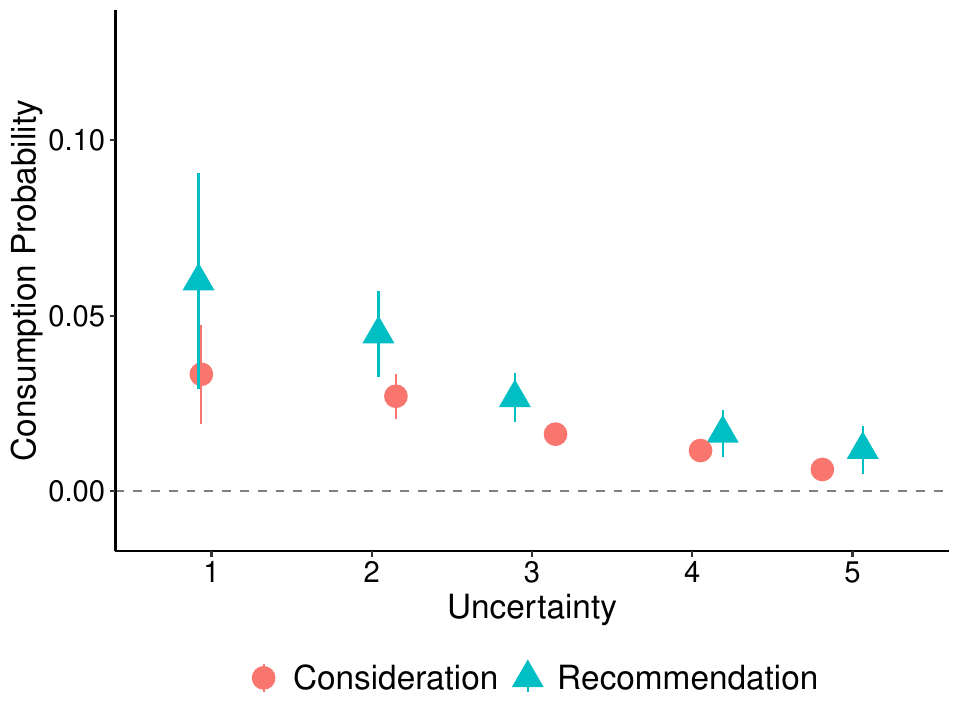}
        \caption{Uncertainty}
    \end{subfigure}
    \begin{subfigure}{.45\textwidth}
        \includegraphics[width=1\linewidth]{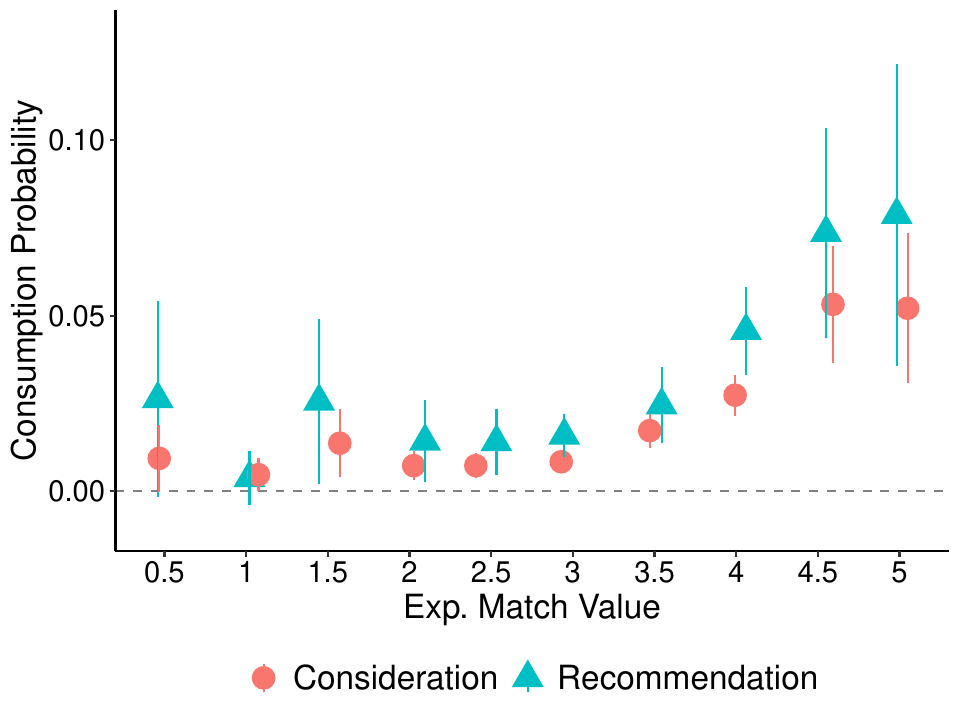}
        \caption{Expected Match Value}
    \end{subfigure}
    \begin{minipage}{1\linewidth}
        \vspace*{.5em}
        \caption{Beliefs Explain Consumption: Heterogeneous Effect by Treatment}
        \label{figure:reg-consumption-on-rec-hte}
        \vspace*{.5em}
        \small
        \setstretch{1} \emph{Notes}: The figure exhibits the estimates of consumption probability by treatment -- consideration-only set (red circle) and recommendation set (blue triangle) -- and conditional on the reported prior uncertainty (Panel (a)) and the prior expected match value (Panel (b)).
        Since these estimates are conditional on prior beliefs, only goods about which beliefs were elicited were included.
        Lines represent 95\% confidence intervals with clustered standard errors at the consumer level.
    \end{minipage}
\end{figure}

In \hyref{figure:reg-consumption-on-rec-hte}[Figure], we consider how consumption probability for goods in the consideration and recommendation treatment groups depends on the prior level of uncertainty (panel (a)) and expected match value (panel (b)).
The figure highlights that, no matter which prior beliefs, recommendation always increases consumption probability beyond mere consideration.
Furthermore, it indicates that lower uncertainty and higher expected match value are associated with higher consumption probability.
Finally, \hyref{figure:reg-consumption-on-rec-hte}[Figure] provides suggestive evidence that the effect of expected match value on consumption is virtually unchanged by recommendation (a level change), while recommendations strengthen the relationship between uncertainty and consumption (a steeper relationship).

\setstretch{1}

\begin{figure}[t]
    \centering
    \begin{subfigure}[t]{.45\textwidth}
        \includegraphics[width=1\linewidth]{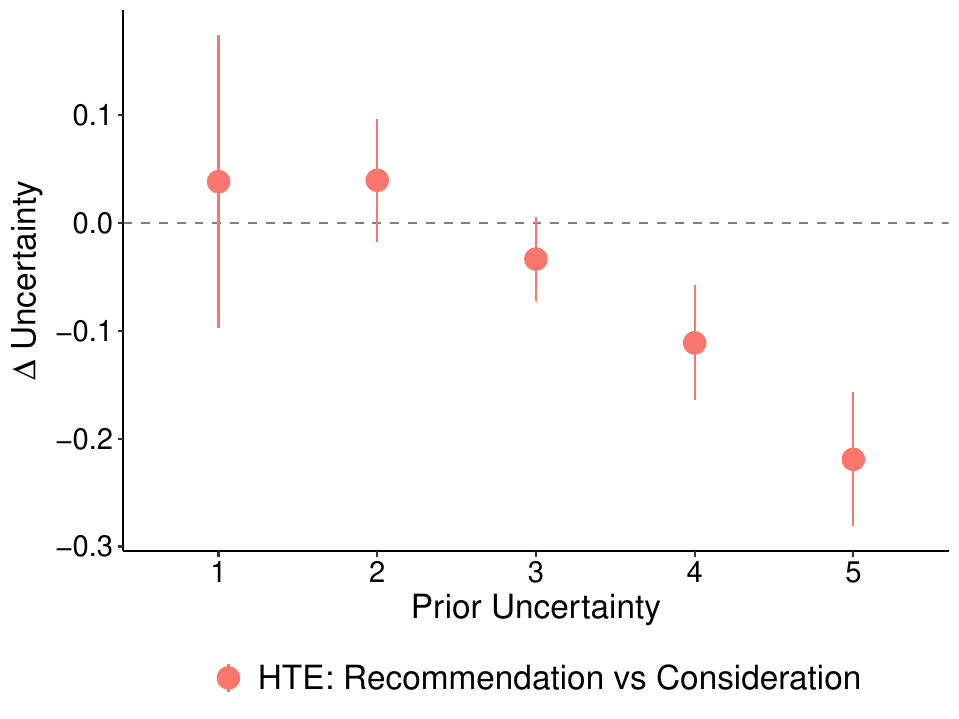}
        \caption{Uncertainty}
    \end{subfigure}
    \begin{subfigure}[t]{.45\textwidth}
        \includegraphics[width=1\linewidth]{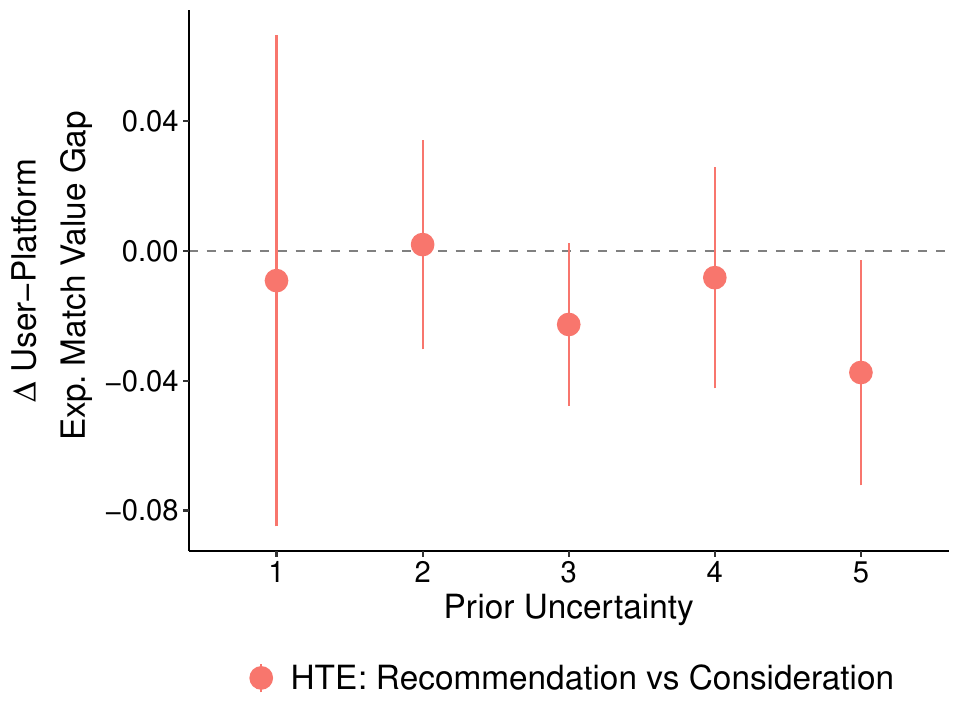}
        \caption{Expected Match Value}
    \end{subfigure}
    \begin{minipage}{1\linewidth}
        \small
        \vspace*{.5em}
        \caption{Impact of Recommendations on Beliefs: Heterogeneous Effect by Treatment}
        \label{figure:reg-beliefs-on-rec-hte}
        \vspace*{.5em}
        \small
        \setstretch{1} \emph{Notes}: The figure estimates the average treatment effect of recommendations on beliefs compared to that of consideration (\hyref{equation:rec-on-beliefs}[Equation]), conditional on prior uncertainty.
        Panel (a) exhibits the estimated treatment effect of recommendations (compared to consideration) on uncertainty, whilst Panel (b) displays the treatment effect on the distance between the consumer's expected value and the platform's predicted value, both conditional on the reported prior uncertainty. 
        Lines represent 95\% confidence intervals with clustered standard errors at the consumer level.
        The change is taken to be over the course of the experiment, considering the first and last beliefs reported for each consumer $i$ and good $x$.
    \end{minipage}
\end{figure}
\setstretch{1.3}

In order to better understand the heterogeneity based on prior beliefs, we assess if and to what extent consumers differentially updated their beliefs based on their baseline uncertainty levels. 
To do so, we reestimate \hyref{equation:rec-on-beliefs}[Equation]:
\begin{align}
    \label{equation:rec-on-beliefs-heterogeneity}
    y_{i,x,t} &= \beta_0 + \beta_1 r_{i,x} + \epsilon_{i,x} \tag{4}
\end{align}
again, where we consider $y_{i,x,t} = \Delta \sigma_{i,x}^b$ and $y_{i,x,t} = \Delta |v_{i,x}^p-v_{i,x}^b|$. 
Differently from before, we estimate this specification \emph{conditional on prior beliefs}, that is, for each level of prior uncertainty.

The results are presented in \hyref{figure:reg-beliefs-on-rec-hte}[Figure], in panel (a), for the change in uncertainty ($\Delta \sigma_{i,x}^b$), and, in panel (b), for the change in the expected match value gap ($\Delta |v_{i,x}^p-v_{i,x}^b|$).
\hyref{figure:reg-beliefs-on-rec-hte}[Figure][(a)] shows that the information provided by recommendation is largely driven by initially high prior uncertainty and that there is little change in uncertainty for elicitations with initially low uncertainty levels. 
This is consistent with our theoretical framework and intuitively plausible as there is more scope for recommendations to provide informational value the larger the initial uncertainty levels are. 
In line with this observation, \hyref{figure:reg-beliefs-on-rec-hte}[Figure][(b)] indicates that the expected match value gap is brought closer to the platform's predicted value the higher prior uncertainty is, but overall the estimates are less precise.

Overall, combined with the results of \hyref{figure:reg-consumption-on-rec-hte}[Figure], the data suggest that the effects of recommendation are moderated by the degree of prior uncertainty, and thus provide further evidence of the informational role played by recommendations.

\subsubsection{Consumer Experience}
A related dimension of potential heterogeneity in the effects of recommendations is that of consumers' past experience. 
Our underlying working hypothesis is that consumers who have explored a significant portion of the product space are likely to hold more precise beliefs about the match values of goods. 
Indeed, as mentioned earlier, we find a negative association between prior uncertainty and previous consumer experience (see column (3) of \hyref{table:reg-beliefs-on-popularity}[Table]). 
Consequently, the informational gain from recommendations may be lower for such more experienced consumers. 
We proxy consumer experience using the log of past consumption, specifically the number of movies rated prior to the experimental intervention.

To assess whether the extent of past consumption experience mediates how recommendations impact consumption, we interact the recommendation term in specification \hyref{equation:direct-rec-on-consumption1}[Equation] with our measure of consumer experience. 
The results, presented in \hyref{table:reg-consumption-on-rec-hte-experience}[Table], use the same specifications considered in \hyref{figure:reg-consumption-on-rec}[Table]. 
The estimates indicate a negative heterogeneous effect: recommendations have less of an effect on consumption for more experienced consumers, with recommendations having approximately 20\% less impact on consumption for consumers who had watched 500 movies at the outset of the experiment compared to those who had watched only 100.

\setstretch{1}
\begin{table}[t]\setstretch{1.1}
	\centering
	\begin{tabular}{l@{\extracolsep{4pt}}ccc@{}}
  \hline\hline
  & \multicolumn{3}{c}{Consumption} \\
  \cline{2-4}
  \hspace*{9em} & (1) & (2) & (3) \\
  \hline
  Consideration & 0.0004$^{**}$ & 0.002$^{**}$ &  \\ 
   & (0.0002) & (0.001) &  \\ [.1em]
  Recommendation & 0.011$^{***}$ & 0.053$^{***}$ & 0.045$^{**}$ \\ 
   & (0.003) & (0.012) & (0.019) \\ [.1em]
  Recommendation $\times$ Log(Past Consumption) & $-$0.001$^{*}$ & $-$0.006$^{***}$ & $-$0.005$^{*}$ \\ 
   & (0.0005) & (0.002) & (0.003) \\ [.1em]
  Log(Past Consumption) & $-$0.0001 & $-$0.001$^{*}$ & $-$0.0004 \\ 
   & (0.0003) & (0.001) & (0.001) \\ [.1em]
  Constant & 0.005$^{**}$ & 0.019$^{***}$ & 0.014$^{*}$ \\ 
   & (0.002) & (0.005) & (0.007) \\ [.1em]
  \hline \\
  Observations & 1,026,342 & 56,040 & 20,715 \\ 
  R$^{2}$ & 0.001 & 0.004 & 0.003 \\ 
  \hline\hline
  \multicolumn{4}{l}{\footnotesize Clustered standard errors at the subject level in parentheses.}\\
  \multicolumn{4}{l}{\footnotesize $^{*}$ \(p<0.1\), $^{**}$ \(p<0.05\), $^{***}$ \(p<0.01\)}
\end{tabular}
    \begin{minipage}{1\textwidth}
        \vspace*{.5em}
        \caption{Heterogeneous Effects of Recommendation by Consumption Experience}
        \label{table:reg-consumption-on-rec-hte-experience}
        \setstretch{1} \emph{Notes}:
        This table tests whether recommendations impact consumption differently depending on consumers' past experience, as proxied by the log of past consumption, that is, the number of movies rated at the outset of the experimental intervention.
        The estimates correspond to those specified in \hyref{equation:direct-rec-on-consumption1}[Equation] expanded to include the log of past consumption and its interaction with whether a good was in the recommendation treatment for a particular consumer.
        Each component displays the baseline control and the estimated average treatment effect of consideration and recommendation on consumption for the different sample specifications. 
    \end{minipage}
\end{table}
\setstretch{1.3}

This conclusion is further supported by considering how persuasion rates change when focusing on consumers with above-median consumption experience at the outset of the intervention. 
While the persuasion rate for the consideration-only treatment remains similar for more experienced consumers (0.64\% compared to 0.73\%), the persuasion rate for recommendations drops significantly, from 33.34\% in the overall sample to 22.35\% for this subset.

In short, recommendations have less sway over consumption for consumers who are more experienced, and thus better informed.
These findings are consistent with a decreasing marginal value of information and support the predominance of an informational channel in the effect of recommendations on consumption in our setting.

\subsection{Assessing Robustness}
\label{section:robustness:robustness}

We now turn to a discussion of several robustness checks to further validate our findings.

\subsubsection{Consumer Fixed Effects} 
The first possible concern is that there is a large amount of consumer heterogeneity -- both in their propensity to consume goods, but also in their frequency of visiting the platform and providing beliefs. 
As such, it is possible that some of the results that we find are due to idiosyncratic differences across consumers and that the effects are driven by a few power consumers. 
In \hyref{online-appendix:tables-fe}, we rule out this explanation by redoing all of the analyses with consumer fixed effects. 
We find that this leads to little difference in the economic magnitude and statistical significance of all the analyses conducted in the main portion of the paper.

\subsubsection{Normalised Beliefs} 
The second possible concern is that, along the lines of consumer heterogeneity, a five-star rating for one consumer has a different interpretation than for another consumer.\footnote{
    This is a canonical issue for interpreting ratings data since before recommender systems used these data extensively (e.g., see \citep{Greenleaf1992JMR}).
} 
We explore if this influences our conclusions regarding the link between recommendations and beliefs. 
To this effect, we produce consumer-specific, normalised measures of uncertainty and expected match values:
For each consumer, we subtract from the elicited specific measure (uncertainty or expected match value) the consumer-specific mean at the first elicitation and divide by the consumer-specific standard deviation. 
We then re-estimate \hyref{equation:rec-on-beliefs}[Equations] using the standardised beliefs. 
We report the results in \hyref{table:reg-beliefs-on-rec-standardised}[Table] and in \hyref{table:reg-beliefs-on-rec-standardised-fe}[Table] including consumer fixed effects -- in \hyperref[online-appendix:additional-tables]{Online Appendices B} and \hyperref[online-appendix:tables-fe]{C}, respectively.
This leads to no change in our conclusions regarding the magnitude of the effects or their economic/statistical significance.

\subsubsection{Timing of Belief Elicitation} 
The final point we address regards the timing of the belief elicitation considered when analysing the causal relationship between beliefs and consumption. 
Specifically, in \hyref{table:iv-consumption-on-beliefs}[Table] we consider only the last elicitation of a given (consumer, good) pair, as it would be the one that is significant for consumption.
However, in \hyref{table:iv-consumption-on-beliefs-1st-elicit}[Table] (\hyref{online-appendix:additional-tables}), we show that our conclusions on the causal impact of beliefs on consumption remain robust to considering the first elicitation, instead of the last.

\section{Discussion}
\label{section:discussion}

In this section, we provide a discussion of the implications for the results as well as relevant caveats given the aspects that we cannot capture with our experimental intervention.

\subsubsection{Welfare Effects} 
Our results suggest that recommendations provide consumers with information that decreases uncertainty about the goods and shifts consumption. 
This does not directly imply that such recommendations are welfare-improving. 
In particular, by decreasing uncertainty, recommendations could be steering consumers toward goods they like less, generating an informational trap either due to risk aversion or by preempting consumers from seeking further information on alternative products.\footnote{
    For a recent paper that makes this point clearly in social learning environments, see \citet{HoezelmannMansoNagarajTranchero2024WP}.
} 
MovieLens, however, does not provide consumption opportunities and deploys recommendation algorithms that are deemed to guide consumers toward ex-post optimal goods.
In short, as recommendations are personalised to consumers and target goods with high match value, their effects on shifting consumption suggest an increase ex-post, realised utility, while the reduction of uncertainty provides an immediate increase in expected payoffs for risk-averse consumers.

\subsubsection{External Validity} 
One potential limitation is whether the findings from this analysis generalise to other domains. 
The intervention was conducted on MovieLens, a platform with two notable features: it lacks profit-driven motives in its recommender system, and it focuses exclusively on a single product type -- movies. 
By contrast, many online platforms are profit-oriented, and recommender systems are deployed across diverse domains, such as news and social media, where consumption processes may differ from those for movies.

Nevertheless, we believe our findings offer meaningful insights for online platforms. 
First, MovieLens data have been instrumental in advancing research on recommendation algorithms, serving as a benchmark for evaluating new algorithms since the 1990s \citep{HarperKonstan2015ACMTIIS}, and inevitably influencing the development of algorithms employed by commercial platforms. 
Second, the collaborative filtering algorithms used by MovieLens are widely applied across domains, including news, social media, and online marketplaces. 
While these contexts involve varying consumption behaviours, the conceptual framework and experimental design in this paper provide tools to disentangle underlying mechanisms and estimate their relative importance in environments driven by collaborative filtering algorithms \citep{EkstrandLudwigKolbRiedl2011}.

In short, our work allows for an evaluation of the mechanisms through which recommendations influence choices on a platform foundational to algorithmic development. 
It serves as a benchmark for assessing the relative magnitudes of economic mechanisms driving choices under a transparent, canonical recommender system, absent profit motives. 
Furthermore, the approach developed here is portable and can be applied to other contexts to deepen our understanding of how collaborative filtering algorithms influence choices under varying objectives, consumption behaviours, and platform constraints.

\subsubsection{Implications for Recommendation System Evaluation} 
A key distinction in the literature on recommendation system evaluation lies in whether the primary goal is to expand consumers' consideration sets or to provide better information about goods they might already consider. 
As mentioned in our discussion of the related literature, both approaches leverage collaborative filtering algorithms to estimate preferences, but they apply different heuristics to generate recommendations. 
Some systems prioritise surfacing items to broaden consumers' options, while others focus on enhancing the informational value of recommendations to provide users with more useful insights. 

Our findings shed light on the relative importance of these channels by showing that recommendations primarily operate through the informational channel rather than by expanding consideration sets. 
This suggests that systems designed to inform consumers -- by helping them make better choices from within their existing options -- are more effective at driving consumption than those focused on consideration set expansion. 

This shift in focus has important implications for recommendation system design. 
As discussed in the introduction, several metrics are often used as proxies for recommendation informativeness, but they conflate broadening options with providing consumers with information, without directly capturing how recommendations achieve the latter. 
In our experimental intervention, we elicited belief data to explicitly measure the informational value of recommendations, demonstrating its potential for optimising recommendation systems. 
In a companion paper \citep{aridor2024movielens}, we adapt our methodology and introduce a procedure for collecting belief data, suitable for large-scale implementation, along with an open-source dataset from MovieLens that incorporates this information. 
Together, these contributions help us better understand the economic forces behind consumption choices in recommendation systems and can be used to guide their design and evaluate their impact.

\section{References}
\vspace*{-0em}\setlength{\bibhang}{0pt}
\bibliographystyle{econ-aea.bst}
{\setstretch{1.15}\setlength{\bibsep}{.0em plus .0ex}
\bibliography{Recommendations-References-All.bib}

\afterpage{\clearpage}
\newpage

\renewcommand{\thesection}{Appendix}
\renewcommand{\thesubsection}{Appendix \Alph{subsection}}
\renewcommand{\thesubsubsection}{\thesubsection.\arabic{subsubsection}}

\titleformat{\subsubsection}[block]
{\normalfont\large\bfseries}{\thesubsubsection.}{.5em}{\large\bfseries}
\titlespacing*{\subsubsection}{0pt}{*1}{*0}

\titleformat{\paragraph}[block]
{\normalfont\large\bfseries}{\theparagraph.}{.5em}{\large\bfseries}
\titlespacing*{\paragraph}{0pt}{*1}{*0}

\section{~}

\subsection{Belief Data Validation Exercises}
\label{appendix:beliefs-data}
In this appendix, we demonstrate that the belief data we collect not only exhibit reasonable patterns, but are also informative about the resulting match value. In other words, we provide evidence that consumers have well-formed beliefs about films and that survey-based measures can accurately capture them.

\begin{table}[b!]\setstretch{1.1}
    \centering
    \begin{tabular}{l@{\extracolsep{4pt}}ccc@{}}
  \hline\hline
  & Realised Match Value & Prior Consumer-Platform  & Prior Uncertainty \\
  &                      & Exp. Match Value Gap & \\
  \cline{2-4}
  \hspace*{9em} & (1) & (2) & (3) \\
  \hline
  Exp. Match Value & 1.017$^{***}$ &  &  \\ 
   & (0.010) &  &  \\ [.1em]
  Uncertainty &  & 0.073$^{***}$ &  \\ 
   &  & (0.023) &  \\ [.1em]
  Log(Past Consumption) &  &  & $-$0.465$^{***}$ \\ 
   &  &  & (0.060) \\ [.1em]
  Log(Popularity) &  &  & $-$0.056$^{*}$ \\ 
   &  &  & (0.033) \\ [.1em]
  Film is Sequel &  &  & $-$0.061$^{***}$ \\ 
   &  &  & (0.010) \\ [.1em]
  Constant &  & 1.128$^{***}$ & 4.167$^{***}$ \\ 
   &  & (0.089) & (0.227) \\ [.1em]
  \hline \\
  Observations & 408 & 21,283 & 20,788 \\ 
  R$^{2}$ & 0.460 & 0.008 & 0.020 \\ 
  \hline\hline
  \multicolumn{4}{l}{\footnotesize Clustered standard errors at the subject level in parentheses.}\\
  \multicolumn{4}{l}{\footnotesize $^{*}$ \(p<0.1\), $^{**}$ \(p<0.05\), $^{***}$ \(p<0.01\)}
\end{tabular}
    \begin{minipage}{1\textwidth}
        \vspace*{.5em}
        \caption{Properties of Belief Data}
        \label{table:reg-beliefs-on-popularity}
        \setstretch{1} \emph{Notes}: This table demonstrates the sensible patterns in the belief data. 
        Column (1) estimates the correlation between realised match value and expected match value. 
        Column (2) estimates the relationship between the distance between expected and realised match value on prior consumer uncertainty. 
        Column (3) displays the relationship between consumer uncertainty and 
        consumers' log of past consumption as given by the number of films rated at the outset of the experimental period, 
        films' log popularity as given by the number of ratings at the outset of the experimental period on the MovieLens platform, 
        and whether the film is a sequel or not.
        Prior beliefs (uncertainty and expected match value) refer to the first belief elicitation about a particular film from a particular consumer.
    \end{minipage}
\end{table}

First, we show that consumers' beliefs are an \emph{unbiased} statistic for their value assessments after consumption, arguably settling any question about the validity of the expected match value measure. We estimate:
\[
v_{i,x} = \beta_1 v_{i,x}^b + \epsilon_{i,x}
\]
where \(v_{i,x}\) denotes the realised match value of good \(x\) for consumer \(i\), and, in order to capture the initial beliefs of consumers before any intervention, \(v_{i,x}^b\) denotes the consumers' first belief elicitation about good \(x\) for consumer \(i\). The results, in column (1) of \hyref{table:reg-beliefs-on-popularity}[Table], show the estimated coefficient \(\beta_1\) is a precisely estimated 1: prior beliefs of consumers are on average correct.

Second, we show that the (Euclidean) distance between the expected match value assessment and the realised match value is increasing in the reported uncertainty level. Specifically, we estimate:
\[
\left|v_{i,x} - v_{i,x}^b\right| = \beta_0 + \beta_1 \sigma_{i,x}^b + \epsilon_{i,x}.
\]
Column (2) of \hyref{table:reg-beliefs-on-popularity}[Table] reports \(\beta_1 > 0\), a positive relationship between uncertainty and the resulting difference, validating that larger uncertainty results in less aligned belief and actual match value.

We then explore the relationship between expected match value and uncertainty and show in  \hyref{figure:beliefs-uncertainty-by-expvalue}[Figure] that, as one would expect, consumers are more certain of extreme value assessments (i.e., close to 0 or 5 stars) relative to more moderate ones (i.e., 3 stars). 

The final validity check that we conduct explores how uncertainty relates to the popularity of a good (measured by the number of community ratings on MovieLens), the past consumption experience of a consumer (measured by their number of pre-experiment ratings), and whether the film is a sequel or not (measured by joining to IMDb data). We therefore run the following regression to assess the association between consumers' reported uncertainty and these measures:
\begin{equation}
\sigma_{i,x}^b = \beta_1 \log(\textrm{Past Consumption}_{i}) + \beta_2 \log(\text{Popularity}_x) + \beta_3 \text{Sequel}_x + \epsilon_{i,x}
\end{equation}
where the notation is similar to the previous specifications. To isolate the possible role of the experimental intervention in modifying beliefs, we restrict focus to the first belief elicitation of a given good for each consumer.

Column (3) of \hyref{table:reg-beliefs-on-popularity}[Table] displays the results, showing that greater popularity and being a sequel are associated with lower uncertainty. Interestingly, it also shows that uncertainty is decreasing in past consumption -- indicating that less experienced consumers are, on average, more uncertain about their expected match value for goods.

\begin{figure}[t]\setstretch{1.1}
    \centering
    \includegraphics[width=.5\linewidth]{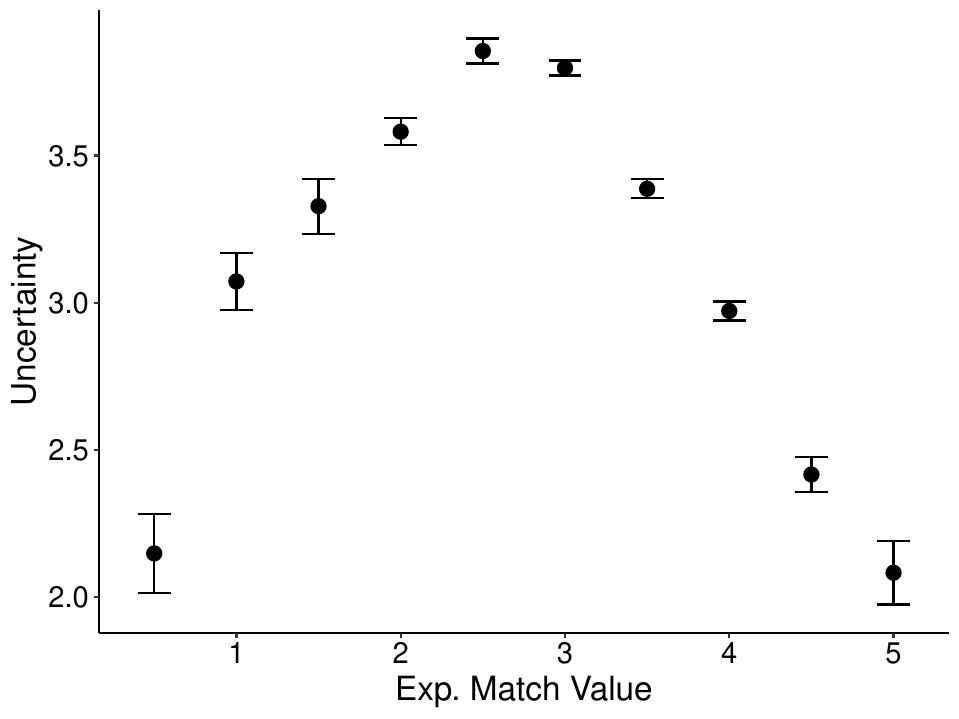}
    \begin{minipage}{1\textwidth}
        \vspace*{.5em}
        \caption{Uncertainty by Expected Quality}
        \label{figure:beliefs-uncertainty-by-expvalue}\setstretch{1} \emph{Notes}: This figure shows the expected match value on the x-axis and the associated conditional average uncertainty score on the y-axis, as well as the associated 95\% confidence interval.
    \end{minipage}
\end{figure}

\begin{table}[t]\setstretch{1.1}
    \centering
    \begin{tabular}{l@{\extracolsep{4pt}}ccccccc@{}}
  \hline\hline
 & Mean & Std. Dev. & \multicolumn{5}{c}{Percentile} \\ 
 &  &  & Median & 25th & 75th & Min & Max \\ 
 \cline{2-8}
 \hspace*{9em} & (1) & (2) & (3) & (4) & (5) & (6) & (7) \\
  \hline
  Realised Match Values & 3.71 & 0.88 & 4.00 & 3.50 & 4.00 & 0.50 & 5.00 \\ [.1em]
  \hspace*{1em}\emph{Avg. per user} & 3.77 & 0.59 & 3.83 & 3.50 & 4.16 & 0.50 & 5.00 \\ [.5em]
  Exp. Match Values & 3.02 & 0.94 & 3.00 & 2.50 & 3.50 & 0.50 & 5.00 \\ [.1em]
  \hspace*{1em}\emph{Avg. per user} & 3.02 & 0.57 & 3.05 & 2.70 & 3.40 & 0.50 & 4.53 \\ [.4em]
  Uncertainty & 3.40 & 1.15 & 3.00 & 3.00 & 4.00 & 1.00 & 5.00 \\ [.1em]
  \hspace*{1em}\emph{Avg. per user} & 3.38 & 0.71 & 3.33 & 2.90 & 3.88 & 1.00 & 5.00 \\ [.5em]
  Platform Exp. Match Values & 3.98 & 0.50 & 4.02 & 3.71 & 4.31 & 0.00 & 5.00 \\ [.1em]
  \hspace*{1em}\emph{Avg. per user} & 3.98 & 0.43 & 4.01 & 3.72 & 4.27 & 1.58 & 5.00 \\ [.5em]
  Platform Exp. Match Values & 4.36 & 0.42 & 4.37 & 4.09 & 4.69 & 2.64 & 5.00 \\ 
  for Recommended &&&&&&& \\
  [.1em]
  \hspace*{1em}\emph{Avg. per user} & 4.39 & 0.37 & 4.42 & 4.13 & 4.68 & 2.81 & 5.00 \\ 
  \hline\hline
\end{tabular}
    \begin{minipage}{1\textwidth}
        \vspace*{.5em}
        \caption{Belief Data Summary Statistics}
        \label{table:summary-stats-beliefs}
        \setstretch{1} \emph{Notes}: We present summary statistics on the belief data. We show the distributional statistics for the realised and expected match values, the uncertainty measures, the platform's predicted match values for the films considered in the intervention, and the platform's predicted match value for the set of goods that show up in recommendations. Within each field, the first row shows the distribution across the full dataset, whereas the second row shows the distribution across consumers (averaging to a single average value for each consumer).
    \end{minipage}
\end{table}

We conclude the validation section by presenting summary statistics about the belief data and the experiment, presented in \hyref{table:summary-stats-beliefs}[Table]. \hyref{table:summary-stats-beliefs}[Table] highlights the differences between the realised and expected match values, the uncertainty measures, and the platform's predicted match values for the full set of goods in the experimental intervention versus the goods that show up in recommendations. Overall, we collect 39,608 different belief elicitations for 6,674 distinct films and 1,031 distinct goods. During the experimental period, 8,902 films were rated, with 6,240 qualifying as consumption under our robust measure. \hyref{table:summary-stats}[Table] additionally provides summary statistics of the number of completed belief surveys and the number of films consumed per consumer.

\begin{table}[t]\setstretch{1.1}
    \centering
    \begin{tabular}{l@{\extracolsep{4pt}}ccccccc@{}}
  \hline\hline
  & Mean & Std. Dev. & \multicolumn{5}{c}{Percentile} \\ 
  &  &  & Median & 25th & 75th & Min & Max \\ 
  \cline{2-8}
  \hspace*{9em} & (1) & (2) & (3) & (4) & (5) & (6) & (7) \\
  \hline
  Films Consumed per user & 4.30 & 9.30 & 1.00 & 0.00 & 5.00 & 0.00 & 157.00 \\ [.1em]
  Belief Surveys per user & 3.98 & 5.88 & 2.00 & 1.00 & 4.00 & 1.00 & 72.00 \\ 
  \hline\hline
\end{tabular}
    \begin{minipage}{1\textwidth}
        \vspace*{.5em}
        \caption{Summary Statistics on Consumption and Beliefs}
        \label{table:summary-stats}
        \setstretch{1} \emph{Notes}: This table provides summary statistics on the number of completed belief surveys (each including 10 goods) and the number of films consumed per consumer.
    \end{minipage}
\end{table}

\clearpage

\subsection{Recruitment and Study Implementation: Additional Details}
\label{appendix:recruitment}

In our intervention, we target a random sample from a subset of the platform's users.\footnote{
    The use of or access to the platform is prohibited to individuals under the age of 18, as per the platform's terms of service.
}
In order to mitigate the heterogeneity of treatment effects across consumers arising from differences in the match value of the recommendations, 
we restrict our sample to users who satisfy the following conditions:
\emph{(i)} having rated more than 100 films in total;
\emph{(ii)} having rated fewer than 3,000 films in total; and
\emph{(iii)} over the previous $m=1,2,3,4$ months, having rated a minimum of $\lceil 1.5 m\rceil$ films.
The first condition is a minimum data requirement so that the recommender system algorithm utilised by the platform is able to provide valuable recommendations.
This is especially important given that, throughout the duration of the intervention, the assignment of films to treatments is held fixed and therefore so is the set of films that can be recommended.
The second excludes power users. The last condition seeks to guarantee that the targeted user is minimally active on the platform over the recent past.
These criteria were chosen in consultation with the platform's experts in order to ensure that the data are representative of the overall platform population, with stable users who are familiar with the platform's recommender system.

\subsection{Additional Related Work Discussion}\label{online_appendix:additional_related_work}

We provide some additional discussion of related work for similar, yet distinct, marketing mechanisms such as advertising and online reviews.

\noindent \textbf{Advertising.} 
There is a vast literature on the economics of advertising that similarly attempts to decompose whether and why advertisements influence consumption. 
For example, some \citep[e.g.][]{HonkaHortacsuVitorino2017RAND,TsaiHonka2021MktSc,UrsuSimonovAn2023MktSc} argue that advertising in several contexts acts through the consideration channel, and others \citep[e.g.][]{GrossmanShapiro1984REStud,MeurerStahl1994IJIO,Ackerberg2003IER,SahniNair2020REStud} focus on its informational effects. 
The nuances in measuring the effectiveness and the mechanisms behind the impact of advertising, as discussed in \citet{LewisRao2015QJE}, motivate our empirical design for measuring the impact of recommendations. 
While there are similarities in the underlying mechanisms, advertising and recommendation systems target distinct aspects of the consumer choice process and are generated using different methods and by agents with differing incentives. 
Recommendation systems aggregate consumer data on a platform to provide predictions for multiple goods present on the platform, while advertisements are aimed at persuading consumers to purchase one good or goods from one brand. 
Thus, we view one of our contributions as, similarly to the literature on advertising, decomposing the role of the mechanisms that drive the role of this distinct and relatively new tool for online platforms.

\noindent \textbf{Online Reviews.} 
Another source of possible information that guides consumers in their choices comes from online reviews \citep{Tadelis2016ARE}. 
The basis for most recommendation systems, and the one used in this context, is the set of consumer-provided ratings and reviews that they leave on the platform. 
The primary difference between the two channels is that while reviews provide non-personalised information, a recommendation system provides \textit{personalised} recommendations tailored to the tastes of the consumer. 
Furthermore, the economic mechanisms behind reviews are primarily informational and lack a consideration component since consumers have to actively seek out the good in order to view its reviews. 
Consistent with this, the literature on consumer reviews has primarily focused on issues such as differentiating between the effectiveness of online versus professional reviews \citep{ReimersWaldfogel2021AER} and the feedback loops between consumption and rating \citep{AcemogluMakhdoumiMalekianOzdaglar2022Ecta,BondiRossiStevens2022WP}. 
In our setting, there are no reviews with consumer text -- only ratings. Recommendations, in contrast, are not only distinct from reviews but are also compelling to study in their own right, as they involve complex personalisation algorithms and economic mechanisms that influence consumer choice in ways fundamentally different from reviews. 
While reviews are typically complementary to recommendations, they have distinctly different roles in the consumer choice process.

\newpage

\afterpage{\clearpage}
\newpage

\renewcommand{\thesection}{Online Appendix}
\renewcommand{\thesubsection}{Online Appendix \Alph{subsection}}
\renewcommand{\thesubsubsection}{\Alph{subsection}.\arabic{subsubsection}}
\renewcommand{\theparagraph}{\Alph{subsection}.\arabic{subsubsection}.\arabic{paragraph}}

\titleformat{\subsubsection}[block]
{\normalfont\large\bfseries}{\thesubsubsection.}{.5em}{\large\bfseries}
\titlespacing*{\subsubsection}{0pt}{*1}{*0}

\titleformat{\paragraph}[block]
{\normalfont\large\bfseries}{\theparagraph.}{.5em}{\large\bfseries}
\titlespacing*{\paragraph}{0pt}{*1}{*0}

\section{~}

\subsection{Additional Tables}
\label{online-appendix:additional-tables}

\singlespacing

\begin{table}[th!]\setstretch{1.1}
	\centering
	\begin{tabular}{l@{\extracolsep{4pt}}ccc@{}}
  \hline\hline
  \hspace*{9em} & \multicolumn{3}{c}{Consumption} \\
  \cline{2-4}
  & (1) & (2) & (3) \\
  \hline
  Consideration & 0.0003$^{*}$ & 0.002$^{**}$ &  \\ 
   & (0.0002) & (0.001) &  \\ [.1em]
  Recommendation & 0.005$^{***}$ & 0.013$^{***}$ & 0.015$^{***}$ \\ 
   & (0.0005) & (0.002) & (0.003) \\ [.1em]
  Constant & 0.005$^{***}$ & 0.010$^{***}$ & 0.011$^{***}$ \\ 
   & (0.0003) & (0.001) & (0.001) \\ [.1em]
  \hline 
  Observations & 1,083,123 & 57,396 & 21,208 \\ 
  R$^{2}$ & 0.001 & 0.003 & 0.003 \\ 
  \hline\hline
  \multicolumn{4}{l}{\footnotesize Clustered standard errors at the subject level in parentheses.}\\
  \multicolumn{4}{l}{\footnotesize $^{*}$ \(p<0.1\), $^{**}$ \(p<0.05\), $^{***}$ \(p<0.01\)}
\end{tabular}
    \begin{minipage}{1\textwidth}
        \vspace*{.5em}
        \caption{The Impact of Recommendation on Consumption (\hyref{hypothesis:consumption-on-rec}[Hypothesis])}
        \label{table:reg-consumption-on-rec}
        \setstretch{1} \emph{Notes}:
        This table tests whether consideration and recommendation impact consumption probability.
        The estimates correspond to those specified in \hyref{equation:direct-rec-on-consumption1}[Equation] providing a direct counterpart to \hyref{figure:reg-consumption-on-rec}[Figure].
        Each component displays the baseline control and the estimated average treatment effect of consideration and recommendation on consumption for the different sample specifications. 
        Each observation corresponds to a pair (consumer $i$, good $x$). 
        Column (1) includes the ``full data'', that is, it includes all consumers $i$ and all goods $x$ in the consumer-specific control, exposure-only, and recommendation sets. 
        Column (2) includes only ``considered strata'', i.e., it includes the goods to which a consumer was exposed through the belief elicitation survey, and all the goods in the same consumer-specific stratum. 
        Column (3) further restricts the sample to only ``considered items'', corresponding to goods to which a consumer was exposed through the belief elicitation survey; it necessarily excludes goods in the control set.
    \end{minipage}
\end{table}

\begin{table}[th!]\setstretch{1.1}
	\centering
	\begin{tabular}{l@{\extracolsep{4pt}}ccc@{}}
  \hline\hline
  & \multicolumn{3}{c}{Consumption} \\
  \cline{2-4}
  \hspace*{9em} & (1) & (2) & (3) \\
  \hline
  Uncertainty & $-$0.159$^{**}$ &  & $-$0.162$^{**}$ \\ 
  & (0.067) &  & (0.075) \\ [.1em]
  Exp. Match Value &  & 0.062$^{***}$ & 0.091$^{**}$ \\ 
  &  & (0.019) & (0.037) \\ [.1em]
  Constant & 0.558$^{**}$ & $-$0.171$^{***}$ & 0.293 \\ 
  & (0.229) & (0.057) & (0.226) \\ [.5em]
  \hline \\
  Weak Instruments (Uncertainty) & 7.69 &  & 3.23 \\ 
   & [0.006]  &  & [0.021]  \\ 
  Weak Instruments (Exp. Match Value) &  & 55.19 & 19.82 \\ 
   &  & [0.000]  & [0.000]  \\ 
  Wu-Hausman & 15.71 & 8.76 & 11.92 \\ 
   & [0.000]  & [0.003]  & [0.000]  \\ 
  Sargan &  &  & 0.03 \\ 
   &  &  & [0.858]  \\ 
  Observations & 20,895 & 20,895 & 20,895 \\ 
  \hline\hline
  \multicolumn{4}{l}{\footnotesize Clustered standard errors at the subject level in parentheses.}\\
  \multicolumn{4}{l}{\footnotesize p-values in squared brackets.}\\
  \multicolumn{4}{l}{\footnotesize $^{*}$ \(p<0.1\), $^{**}$ \(p<0.05\), $^{***}$ \(p<0.01\)}
\end{tabular}
    \begin{minipage}{1\textwidth}
        \vspace*{.5em}
        \caption{Beliefs Explain Consumption (\hyref{hypothesis:consumption-on-beliefs}[Hypothesis]): 1st Elicitation}
        \label{table:iv-consumption-on-beliefs-1st-elicit}
        \setstretch{1} \emph{Notes}: 
        This table tests \hyref{hypothesis:consumption-on-beliefs}[Hypothesis] by estimating the causal effect of a good's expected match value and uncertainty on whether it is consumed, as per \hyref{equation:beliefs-on-consumption}[Equation]. 
        It differs from \hyref{table:iv-consumption-on-beliefs}[Table] by considering the first instead of the last time that beliefs about a specific good are elicited from a given consumer.
        The instruments are as described in \hyref{section:mechanisms:consumption-on-beliefs}[Section]; Weak Instruments, Wu-Hausman, and Sargan, correspond to tests regarding weak instruments, endogeneity, and validity of overidentifying restrictions.
    \end{minipage}
\end{table}

\begin{table}[th]\setstretch{1.1}
	\centering
	\begin{tabular}{l@{\extracolsep{4pt}}ccc@{}}
  \hline\hline
  & \multicolumn{3}{c}{Consumption} \\
  \cline{2-4}
  \hspace*{9em} & (1) & (2) & (3) \\
  \hline
  Uncertainty & $-$0.007$^{***}$ &  & $-$0.006$^{***}$ \\ 
   & (0.001) &  & (0.001) \\ [.1em]
  Exp. Match Value &  & 0.010$^{***}$ & 0.010$^{***}$ \\ 
   &  & (0.001) & (0.001) \\ [.1em]
  Constant & 0.038$^{***}$ & $-$0.016$^{***}$ & 0.006 \\ 
   & (0.004) & (0.003) & (0.004) \\ [.1em]
 \hline \\
 Observations & 20,895 & 20,895 & 20,895 \\ 
 R$^{2}$ & 0.004 & 0.006 & 0.009 \\ 
  \hline\hline
  \multicolumn{4}{l}{\footnotesize Clustered standard errors at the subject level in parentheses.}\\
  \multicolumn{4}{l}{\footnotesize $^{*}$ \(p<0.1\), $^{**}$ \(p<0.05\), $^{***}$ \(p<0.01\)}
\end{tabular}
    \begin{minipage}{1\textwidth}
        \vspace*{.5em}
        \caption{Correlation Between Beliefs and Consumption}
        \label{table:reg-consumption-on-beliefs}
        \setstretch{1} \emph{Notes}: 
        This table estimates the correlation between a good's expected match value and uncertainty with its consumption probability as per \hyref{equation:beliefs-on-consumption}[Equation]. 
        It considers only the last time that beliefs about a specific good are elicited from a given consumer. 
        Differently from \hyref{table:iv-consumption-on-beliefs}[Table], the estimates do not have a causal interpretation.
    \end{minipage}
\end{table}

\begin{table}[th]\setstretch{1.1}
	\centering
	\begin{tabular}{l@{\extracolsep{4pt}}ccc@{}}
  \hline\hline
  & \multicolumn{3}{c}{Consumption} \\
  \cline{2-4}
  \hspace*{9em} & (1) & (2) & (3) \\
  \hline
  Uncertainty & $-$0.006$^{***}$ &  & $-$0.006$^{***}$ \\ 
   & (0.001) &  & (0.001) \\ [.1em]
  Exp. Match Value &  & 0.009$^{***}$ & 0.009$^{***}$ \\ 
   &  & (0.001) & (0.001) \\ [.1em]
  Constant & 0.037$^{***}$ & $-$0.013$^{***}$ & 0.008$^{**}$ \\ 
   & (0.004) & (0.003) & (0.004) \\ [.1em]
 \hline \\
 Observations & 20,895 & 20,895 & 20,895 \\ 
 R$^{2}$ & 0.004 & 0.005 & 0.008 \\ 
  \hline\hline
  \multicolumn{4}{l}{\footnotesize Clustered standard errors at the subject level in parentheses.}\\
  \multicolumn{4}{l}{\footnotesize $^{*}$ \(p<0.1\), $^{**}$ \(p<0.05\), $^{***}$ \(p<0.01\)}
\end{tabular}
    \begin{minipage}{1\textwidth}
        \vspace*{.5em}
        \caption{Correlation Between Beliefs and Consumption: 1st Elicitation}
        \label{table:reg-consumption-on-beliefs-1st-elicit}
        \setstretch{1} \emph{Notes}:
        This table estimates the correlation between a good's expected match value and uncertainty with its consumption probability, as per \hyref{equation:beliefs-on-consumption}[Equation].
        It differs from \hyref{table:reg-consumption-on-beliefs}[Table] by considering the first instead of the last time that beliefs about a specific good are elicited from a given consumer. 
        Differently from \hyref{table:iv-consumption-on-beliefs-1st-elicit}[Table], the estimates do not have a causal interpretation.
    \end{minipage}
\end{table}

\begin{table}[th]\setstretch{1.1}
	\centering
	\begin{tabular}{l@{\extracolsep{4pt}}cc@{}}
  \hline\hline
  & $\Delta$ Std. Uncertainty & $\Delta$ Std. Consumer-Platform \\
  &                      & Exp. Match Value Gap \\
  \cline{2-3}
  \hspace*{9em} & (1) & (2) \\
  \hline
  Recommendation & $-$0.067$^{***}$ & $-$0.026$^{**}$ \\ 
   & (0.015) & (0.010) \\ [.1em]
  Constant & $-$0.063$^{***}$ & 0.004 \\ 
   & (0.009) & (0.006) \\ [.1em]
  \hline \\
  Observations & 20,704 & 20,704 \\ 
  R$^{2}$ & 0.001 & 0.0004 \\ 
  \hline\hline
  \multicolumn{3}{l}{\footnotesize Clustered standard errors at the subject level in parentheses.}\\
  \multicolumn{3}{l}{\footnotesize $^{*}$ \(p<0.1\), $^{**}$ \(p<0.05\), $^{***}$ \(p<0.01\)}
\end{tabular}
    \begin{minipage}{1\textwidth}
        \vspace*{.5em}
        \caption{The Impact of Recommendation on Beliefs (\hyref{hypothesis:beliefs-on-rec}[Hypothesis]) -- Standardised Beliefs}
        \label{table:reg-beliefs-on-rec-standardised}
        \setstretch{1} \emph{Notes}:
         Columns (1) and (2) test \hyref{hypothesis:beliefs-on-rec}[Hypothesis][(i)] and \hyref{hypothesis:beliefs-on-rec}[][(ii)], respectively, by estimating the effect of recommendations on the change in uncertainty $\Delta \sigma_{i,x}^b$ and on the change of the absolute difference between the consumer's expected value and the platform's predicted value, $\Delta |v_{i,x}^b-v_{i,x}^p|$, as per \hyref{equation:rec-on-beliefs}[Equation]. 
        It differs from \hyref{table:reg-beliefs-on-rec}[Table] by considering standardised measures of beliefs, subtracting from the elicited specific measure (uncertainty or expected match value) the consumer-specific mean at the first elicitation and dividing by the consumer-specific standard deviation. 
    \end{minipage}
\end{table}

\afterpage{\clearpage}
\newpage

\subsection{Tables with Consumer Fixed Effects}
\label{online-appendix:tables-fe}

\begin{table}[th]\setstretch{1.1}
	\centering
	\begin{tabular}{l@{\extracolsep{4pt}}ccc@{}}
  \hline\hline
  & \multicolumn{3}{c}{Consumption} \\
  \cline{2-4}
  \hspace*{9em} & (1) & (2) & (3) \\
  \hline
  Consideration & 0.0003$^{*}$ & 0.002$^{**}$ &  \\ 
   & (0.0002) & (0.001) & \\ [.1em]
  Recommendation & 0.005$^{***}$ & 0.013$^{***}$ & 0.014$^{***}$ \\ 
   & (0.0005) & (0.002) & (0.002) \\ [.1em]
   Consumer FEs & Yes & Yes & Yes  \\
  \hline \\
  Observations & 1,083,123 & 57,396 & 21,208 \\ 
  R$^{2}$ & 0.029 & 0.044 & 0.067 \\ 
  \hline\hline
  \multicolumn{4}{l}{\footnotesize Clustered standard errors at the subject level in parentheses.}\\
  \multicolumn{4}{l}{\footnotesize $^{*}$ \(p<0.1\), $^{**}$ \(p<0.05\), $^{***}$ \(p<0.01\)}
\end{tabular}
    \begin{minipage}{1\textwidth}
        \vspace*{.5em}
        \caption{The Impact of Recommendation on Consumption (\hyref{hypothesis:consumption-on-rec}[Hypothesis]): Consumer Fixed Effects}
        \label{table:reg-consumption-on-rec-fe}
        \setstretch{1} \emph{Notes}:
        This table tests whether consideration and recommendation impact consumption probability.
        It differs from \hyref{table:reg-consumption-on-rec}[Table] only in that it includes consumer fixed effects.
        Each component displays the baseline control and the estimated average treatment effect of consideration and recommendation on consumption for the different sample specifications. 
        Each observation corresponds to a pair (consumer $i$, good $x$). 
        Column (1) includes the ``full data'', that is, it includes all consumers $i$ and all goods $x$ in the consumer-specific control, exposure-only, and recommendation sets. 
        Column (2) includes only ``considered strata'', i.e., it includes the goods to which a consumer was exposed through the belief elicitation survey, and all the goods in the same consumer-specific stratum. 
        Column (3) further restricts the sample to only ``considered items'', corresponding to goods to which a consumer was exposed through the belief elicitation survey; it necessarily excludes goods in the control set.
    \end{minipage}
\end{table}

\begin{table}[th]\setstretch{1.1}
	\centering
	\begin{tabular}{l@{\extracolsep{4pt}}ccc@{}}
  \hline\hline
  & \multicolumn{3}{c}{Consumption} \\
  \cline{2-4}
  \hspace*{9em} & (1) & (2) & (3) \\
  \hline
  Uncertainty & $-$0.009$^{***}$ &  & $-$0.007$^{***}$ \\ 
   & (0.001) &  & (0.001) \\ [.1em]
  Exp. Match Value &  & 0.013$^{***}$ & 0.011$^{***}$ \\ 
   &  & (0.002) & (0.001) \\ [.1em]
   Consumer FEs & Yes & Yes & Yes  \\
  \hline \\
  Observations & 20,895 & 20,895 & 20,895 \\ 
  R$^{2}$ & 0.070 & 0.072 & 0.074 \\ 
  \hline\hline
  \multicolumn{4}{l}{\footnotesize Clustered standard errors at the subject level in parentheses.}\\
  \multicolumn{4}{l}{\footnotesize $^{*}$ \(p<0.1\), $^{**}$ \(p<0.05\), $^{***}$ \(p<0.01\)}
\end{tabular}
    \begin{minipage}{1\textwidth}
        \vspace*{.5em}
        \caption{Correlation Between Beliefs and Consumption: Consumer Fixed Effects}
        \label{table:reg-consumption-on-beliefs-fe}
        \setstretch{1} \emph{Notes}:
        This table estimates the correlation between a good's expected match value and uncertainty with its consumption probability, as per \hyref{equation:beliefs-on-consumption}[Equation].
        It considers only the first time beliefs about a specific good are elicited from a given consumer.
        It differs from \hyref{table:reg-consumption-on-beliefs}[Table] only in that it includes consumer fixed effects.
    \end{minipage}
\end{table}

\begin{table}[th]\setstretch{1.1}
	\centering
	\begin{tabular}{l@{\extracolsep{4pt}}ccc@{}}
  \hline\hline
  & \multicolumn{3}{c}{Consumption} \\
  \cline{2-4}
  \hspace*{9em} & (1) & (2) & (3) \\
  \hline
  Uncertainty & $-$0.009$^{***}$ &  & $-$0.007$^{***}$ \\ 
   & (0.001) &  & (0.001) \\ [.1em]
  Exp. Match Value &  & 0.012$^{***}$ & 0.010$^{***}$ \\ 
   &  & (0.002) & (0.001) \\ [.1em]
   Consumer FEs & Yes & Yes & Yes  \\
  \hline \\
  Observations & 20,895 & 20,895 & 20,895 \\ 
  R$^{2}$ & 0.070 & 0.070 & 0.073 \\ 
  \hline\hline
  \multicolumn{4}{l}{\footnotesize Clustered standard errors at the subject level in parentheses.}\\
  \multicolumn{4}{l}{\footnotesize $^{*}$ \(p<0.1\), $^{**}$ \(p<0.05\), $^{***}$ \(p<0.01\)}
\end{tabular}
    \begin{minipage}{1\textwidth}
        \vspace*{.5em}
        \caption{Correlation Between Beliefs and Consumption: 1st Elicitation; Consumer Fixed Effects}
        \label{table:reg-consumption-on-beliefs-1st-elicit-fe}
        \setstretch{1} \emph{Notes}:
        This table estimates the correlation between a good's expected match value and uncertainty with its consumption probability, as per \hyref{equation:beliefs-on-consumption}[Equation].
        It differs from \hyref{table:reg-consumption-on-beliefs}[Table] by considering the first instead of the last time that beliefs about a specific good are elicited from a given consumer. 
        It differs from \hyref{table:reg-consumption-on-beliefs-1st-elicit}[Table] only in that it includes consumer fixed effects.
    \end{minipage}
\end{table}

\begin{table}[th]\setstretch{1.1}
	\centering
	\begin{tabular}{l@{\extracolsep{4pt}}cc@{}}
  \hline\hline
  & $\Delta$ Uncertainty & $\Delta$ Consumer-Platform \\
  &                      & Exp. Match Value Gap \\
  \cline{2-3}
  \hspace*{9em} & (1) & (2) \\
  \hline
  Recommendation & $-$0.055$^{***}$ & $-$0.015$^{**}$ \\ 
   & (0.013) & (0.007) \\ [.1em]
   Consumer FEs & Yes & Yes  \\
  \hline \\
  Observations & 21,283 & 21,283 \\ 
  R$^{2}$ & 0.074 & 0.056 \\ 
  \hline\hline
  \multicolumn{3}{l}{\footnotesize Clustered standard errors at the subject level in parentheses.}\\
  \multicolumn{3}{l}{\footnotesize $^{*}$ \(p<0.1\), $^{**}$ \(p<0.05\), $^{***}$ \(p<0.01\)}
\end{tabular}
    \begin{minipage}{1\textwidth}
        \vspace*{.5em}
        \caption{The Impact of Recommendation on Beliefs (\hyref{hypothesis:beliefs-on-rec}[Hypothesis]): Consumer Fixed Effects}
        \label{table:reg-beliefs-on-rec-fe}
        \setstretch{1} \emph{Notes}:
        Columns (1) and (2) test \hyref{hypothesis:beliefs-on-rec}[Hypothesis][(i)] and \hyref{hypothesis:beliefs-on-rec}[][(ii)], respectively, by estimating the effect of recommendations on the change in uncertainty $\Delta \sigma_{i,x}^b$ and on the change of the absolute difference between the consumer's expected value and the platform's predicted value, $\Delta |v_{i,x}^b-v_{i,x}^p|$, as per \hyref{equation:rec-on-beliefs}[Equation]. 
        The change is taken to be over the course of the experiment, considering the first and last beliefs reported for each consumer $i$ and good $x$.
        It differs from \hyref{table:reg-beliefs-on-rec}[Table] only in that it includes consumer fixed effects.
    \end{minipage}
\end{table}

\begin{table}[th]\setstretch{1.1}
	\centering
	\begin{tabular}{l@{\extracolsep{4pt}}cc@{}}
  \hline\hline
  & $\Delta$ Std. Uncertainty & $\Delta$ Std. Consumer-Platform \\
  &                      & Exp. Match Value Gap \\
  \cline{2-3}
  \hspace*{9em} & (1) & (2) \\
  \hline
  Recommendation & $-$0.058$^{***}$ & $-$0.025$^{**}$ \\ 
  & (0.015) & (0.011) \\ [.1em]
  Consumer FEs & Yes & Yes  \\
  \hline \\
  Observations & 20,704 & 20,704 \\ 
  R$^{2}$ & 0.074 & 0.054 \\ 
  \hline\hline
  \multicolumn{3}{l}{\footnotesize Clustered standard errors at the subject level in parentheses.}\\
  \multicolumn{3}{l}{\footnotesize $^{*}$ \(p<0.1\), $^{**}$ \(p<0.05\), $^{***}$ \(p<0.01\)}
\end{tabular}
    \begin{minipage}{1\textwidth}
        \vspace*{.5em}
        \caption{The Impact of Recommendation on Beliefs -- Standardised Beliefs; Consumer FE}
        \label{table:reg-beliefs-on-rec-standardised-fe}
        \setstretch{1} \emph{Notes}:
        This table tests \hyref{hypothesis:consumption-on-beliefs}[Hypothesis] by estimating the causal effect of a good's expected match value and uncertainty on whether it is consumed, as per \hyref{equation:beliefs-on-consumption}[Equation]. 
        Similar to \hyref{table:reg-beliefs-on-rec-standardised}[Table], it employs standardised measures of beliefs, subtracting from the elicited specific measure (uncertainty or expected match value) the consumer-specific mean at the first elicitation and dividing by the consumer-specific standard deviation. 
        It differs from \hyref{table:reg-beliefs-on-rec-standardised}[Table] only in that it includes consumer fixed effects.
    \end{minipage}
\end{table}

\begin{table}[th]\setstretch{1.1}
	\centering
	\begin{tabular}{l@{\extracolsep{4pt}}ccc@{}}
  \hline\hline
  & \multicolumn{3}{c}{Consumption} \\
  \cline{2-4}
  \hspace*{9em} & (1) & (2) & (3) \\
  \hline
  Consideration & 0.0004$^{**}$ & 0.002$^{**}$ &  \\ 
   & (0.0002) & (0.001) &  \\ [.1em]
  Recommendation & 0.011$^{***}$ & 0.053$^{***}$ & 0.041$^{**}$ \\ 
   & (0.003) & (0.012) & (0.018) \\ [.1em]
  Recommendation $\times$ Log(Past Consumption) & $-$0.001$^{*}$ & $-$0.006$^{***}$ & $-$0.004 \\ 
   & (0.0005) & (0.002) & (0.003) \\ [.1em]
   Consumer FEs & Yes & Yes & Yes  \\
  \hline \\
  Observations & 1,026,342 & 56,040 & 20,715 \\ 
  R$^{2}$ & 0.028 & 0.044 & 0.065 \\ 
  \hline\hline
  \multicolumn{4}{l}{\footnotesize Clustered standard errors at the subject level in parentheses.}\\
  \multicolumn{4}{l}{\footnotesize $^{*}$ \(p<0.1\), $^{**}$ \(p<0.05\), $^{***}$ \(p<0.01\)}
\end{tabular}
    \begin{minipage}{1\textwidth}
        \vspace*{.5em}
        \caption{Heterogeneous Effects of Recommendation by Consumption Experience: Consumer Fixed Effects}
        \label{table:reg-consumption-on-rec-hte-experience-fe}
        \setstretch{1} \emph{Notes:} This table tests whether recommendations impact consumption differently depending on consumers' past experience, as proxied by the log of the past consumption, that is, the number of films rated at the outset of the experimental intervention.
        It differs from \hyref{table:reg-consumption-on-rec-hte-experience}[Table] only in that it includes consumer fixed effects.
        The estimates correspond to those specified in \hyref{equation:direct-rec-on-consumption1}[Equation] expanded to also include the log of past consumption and its interaction with whether or not a good was in the recommendation treatment for a particular consumer.
        Each component displays the baseline control and the estimated average treatment effect of consideration and recommendation on consumption for the different sample specifications. 
    \end{minipage}
\end{table}

\afterpage{\clearpage}
\newpage 

\subsection{Screenshots} 
\label{online-appendix:screenshots}

\subsubsection{Platform Interface}

\begin{figure}[h!]
    \centering
    \begin{subfigure}{.8\textwidth}
        \includegraphics[width=1\linewidth]{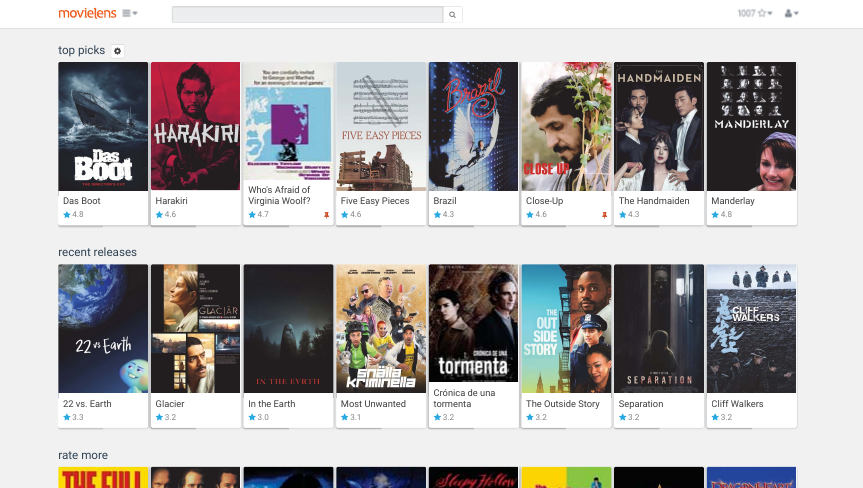}
        \caption{Interface Layout}
        \label{figure:movielens-homepage:layout}
    \end{subfigure}
    \begin{subfigure}{.19\textwidth}
        \includegraphics[width=1\linewidth]{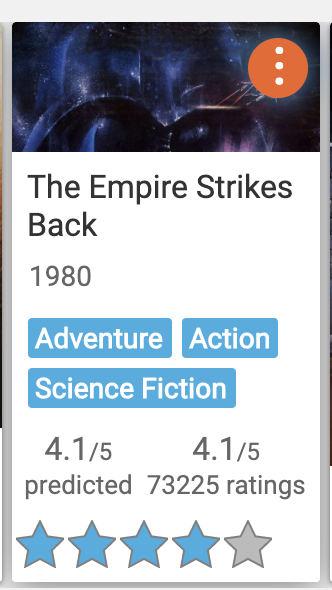}
        \caption{Movie Details Card}
        \label{figure:movielens-homepage:details-card}
    \end{subfigure}
    \begin{minipage}{1\linewidth}
        \small
        \vspace*{.5em}
        \caption{MovieLens Home Page}
        \label{figure:movielens-homepage}
    \end{minipage}
\end{figure}

\begin{figure}[h!]
    \centering
    \includegraphics[width=.8\textwidth]{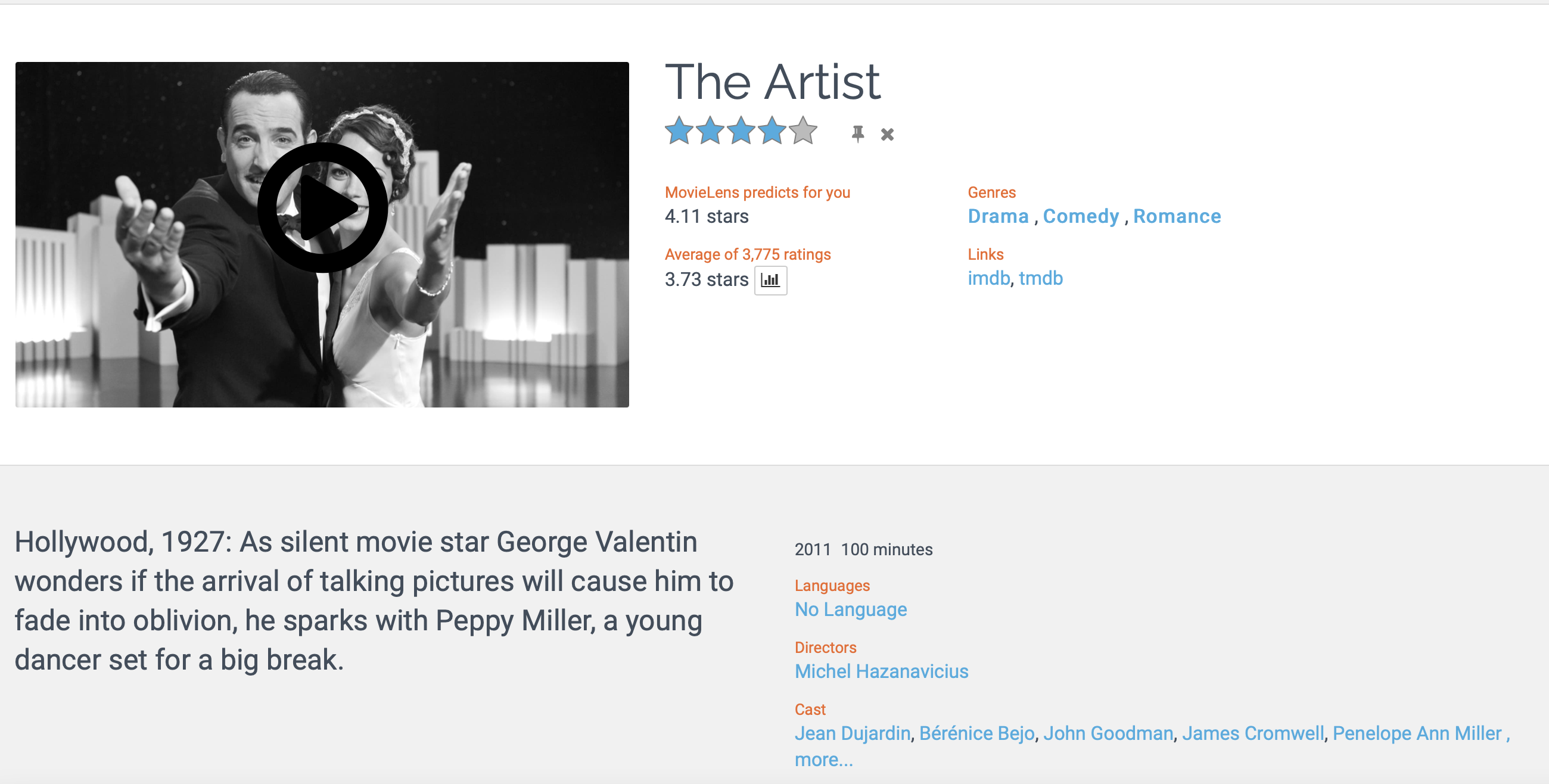}
    \includegraphics[width=.8\textwidth]{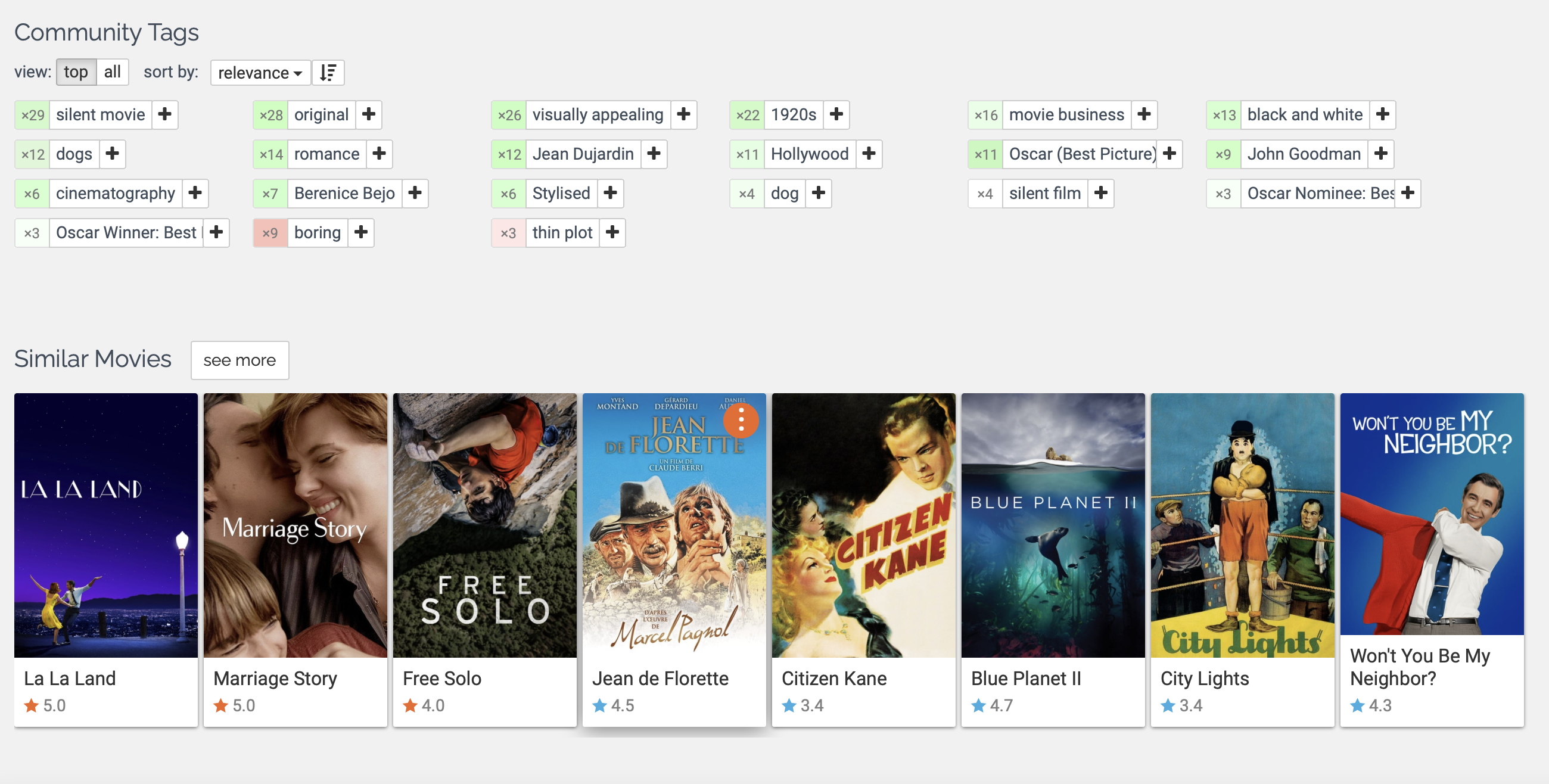}
    \caption{Movie Details Page}
    \label{figure:movielens-movie}
\end{figure}

\afterpage{\clearpage}
\newpage

\subsubsection{Survey}
\label{online-appendix:screenshots:survey}
\begin{figure}[h!]
    \centering
    \includegraphics[width=.8\textwidth]{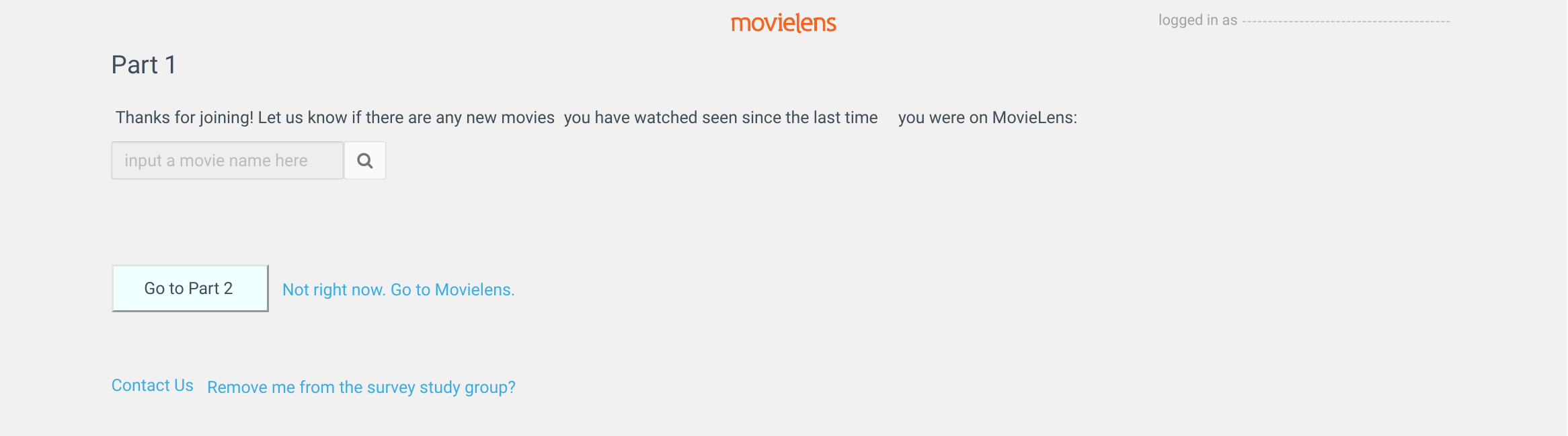}
\end{figure}
\begin{figure}[h!]
    \centering
    \includegraphics[width=.8\textwidth]{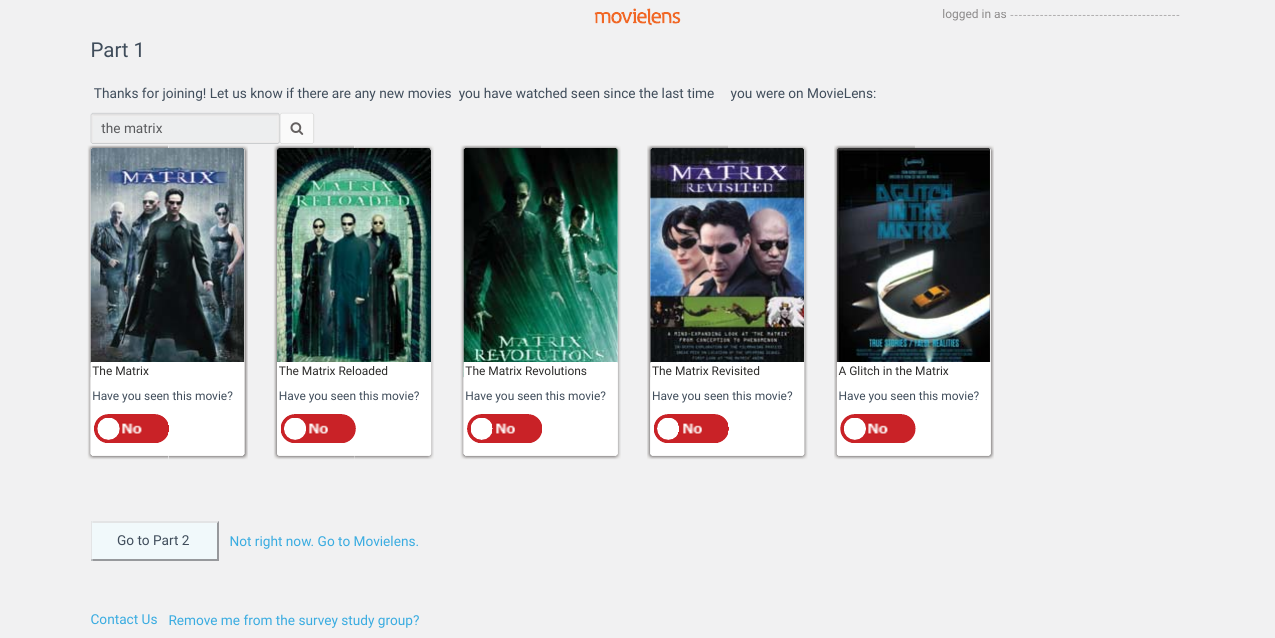}
\end{figure}
\begin{figure}[h!]
    \centering
    \includegraphics[width=.8\textwidth]{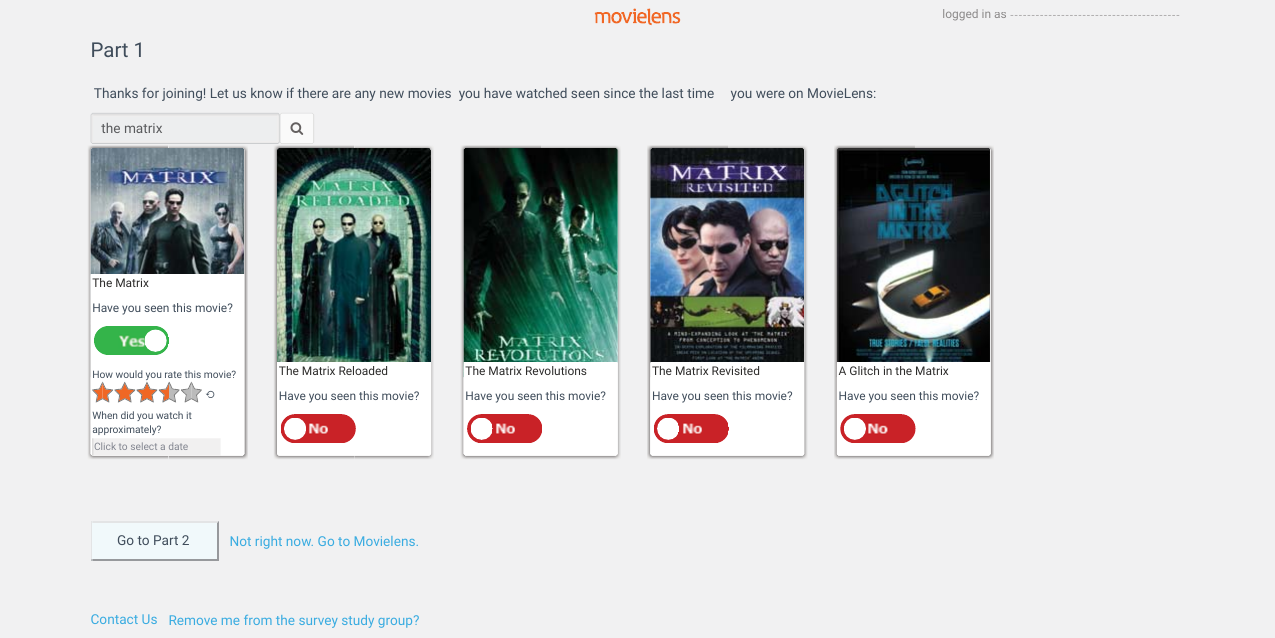}
\end{figure}
% \begin{figure}[h!]
%     \centering
%     \includegraphics[width=.8\textwidth]{Figures/Screenshots/figure-experiment-1.4.pdf}
% \end{figure}
\begin{figure}[h!]
    \centering
    \includegraphics[width=.8\textwidth]{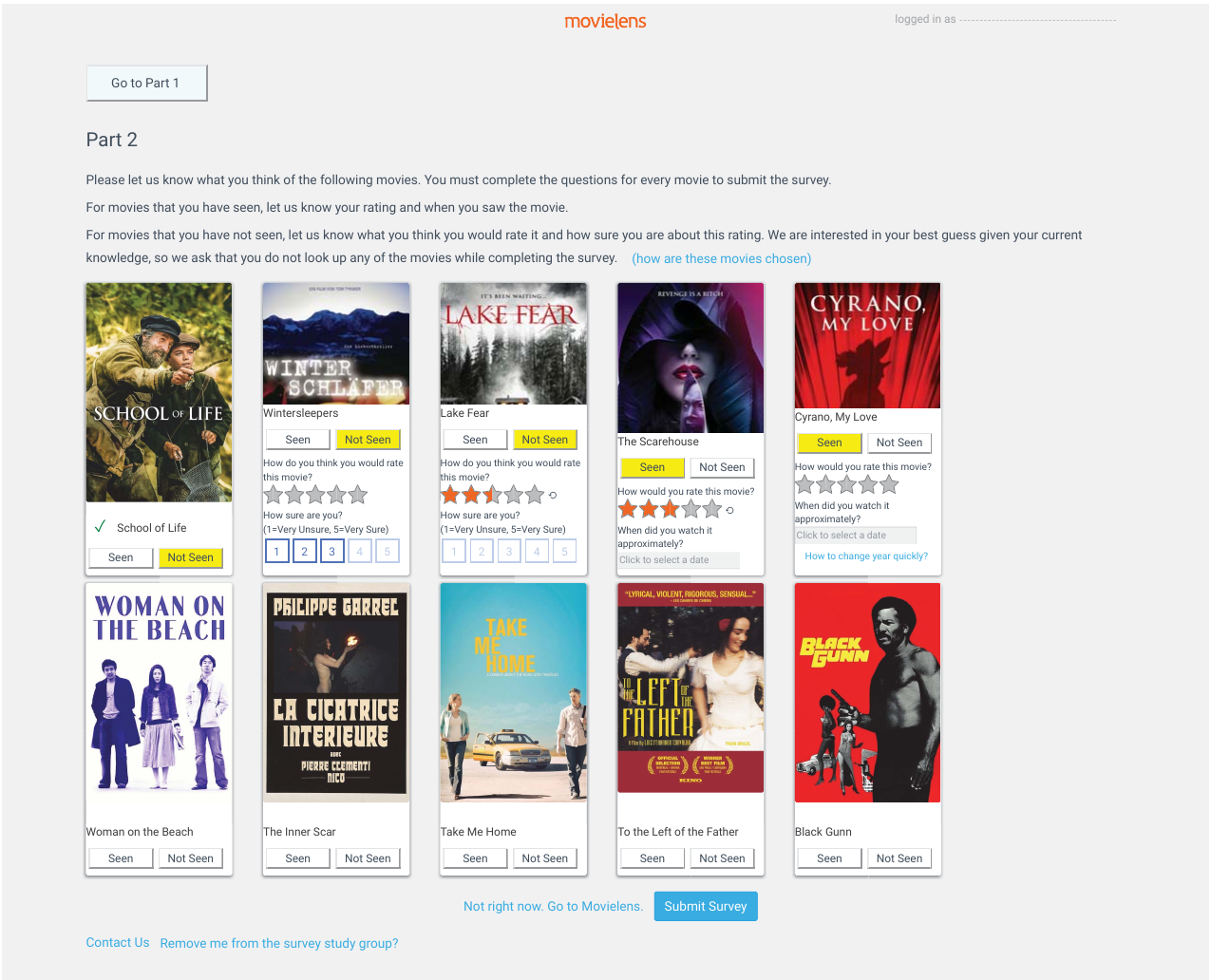}
\end{figure}
\begin{figure}[h!]
    \centering
    \includegraphics[width=.8\textwidth]{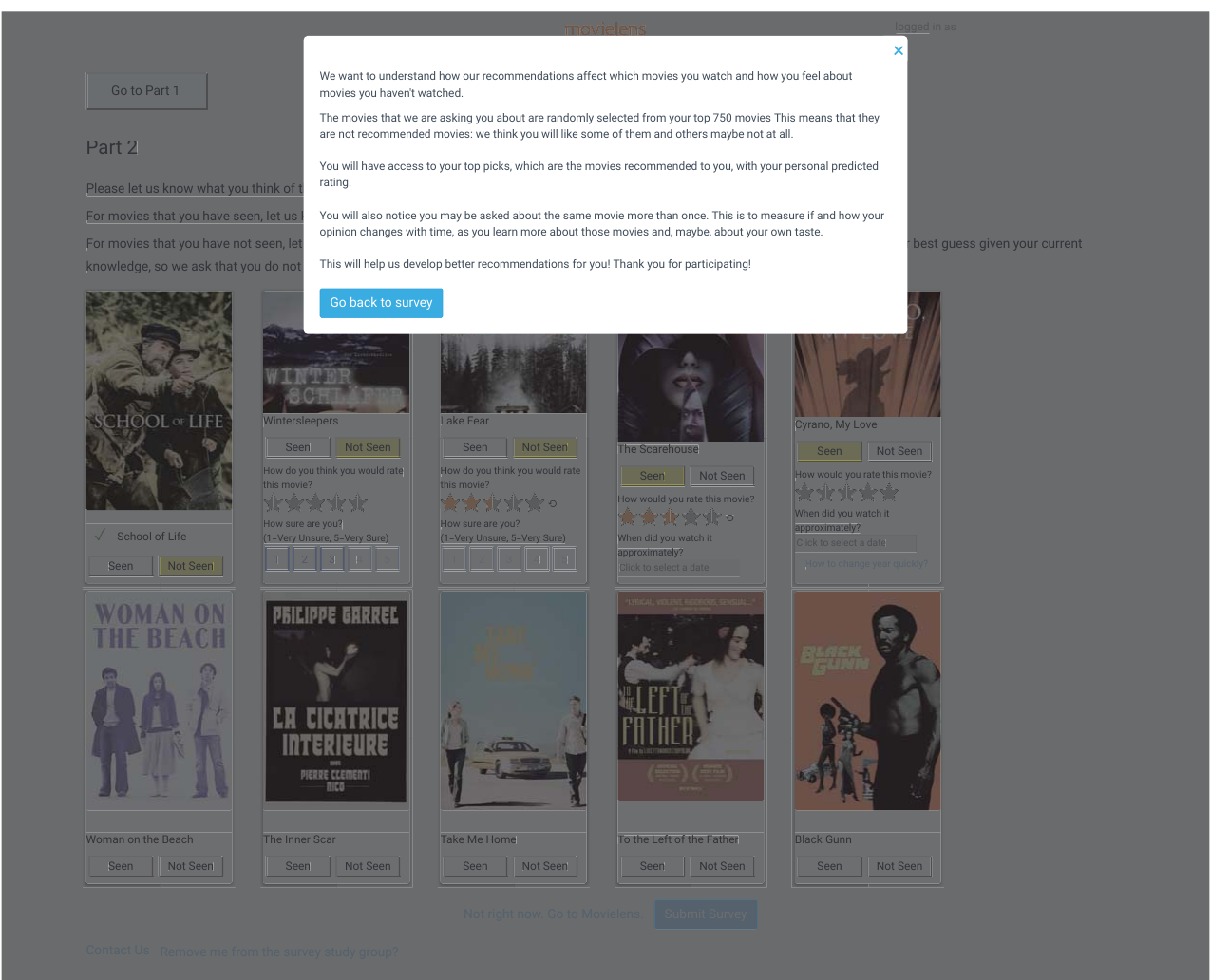}
\end{figure}

\afterpage{\clearpage}
\newpage

\subsubsection{Recruitment}
\label{online-appendix:screenshots:recruitment}
\begin{figure}[h!]
    \centering
    \includegraphics[width=.8\textwidth]{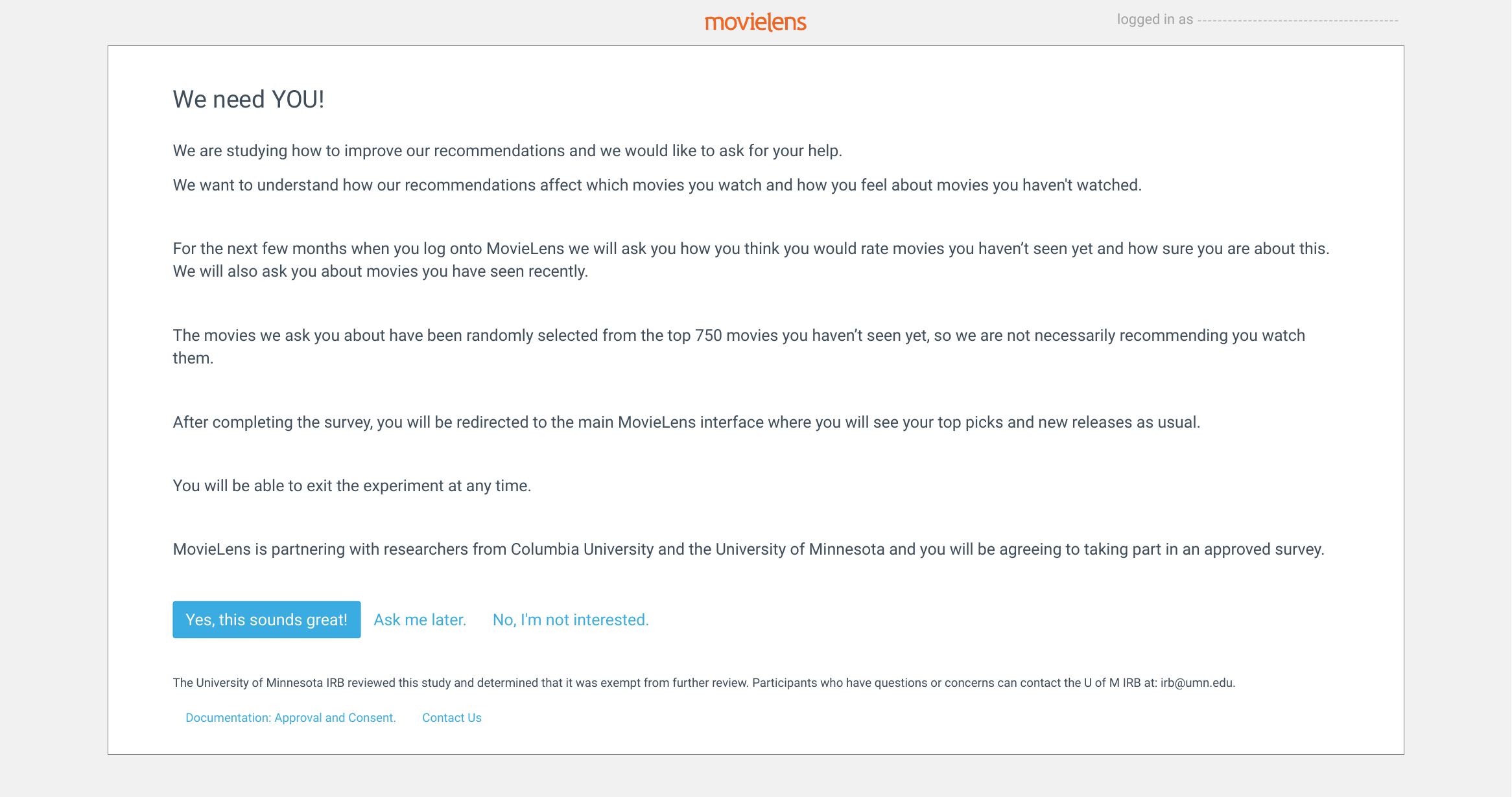}
\end{figure}

\end{document}